\documentclass[a4paper,11pt]{article}
\usepackage[english]{babel}
\usepackage{myjheppub}
\usepackage{subfig}
\usepackage[T1]{fontenc} 
\usepackage{graphicx}
\usepackage{epsfig}
\usepackage{axodraw}
\usepackage{amsmath}
\usepackage{amssymb}
\usepackage{float}
\usepackage{placeins}

\newcommand{\T}{\mathcal{T}}
\newcommand{\D}{\mathcal{D}}
\newcommand{\intd}{\int\,\D^{d}k_{1}\D^{d}k_{2}\,}
\newcommand{\Space}{\qquad\qquad\qquad}

\newcommand{\be}{\begin{equation}}
\newcommand{\ee}{\end{equation}}
\newcommand{\nn}{\nonumber}
\newcommand{\bea}{\begin{eqnarray}}
\newcommand{\eea}{\end{eqnarray}}
\newcommand{\bfig}{\begin{figure}}
\newcommand{\efig}{\end{figure}}
\newcommand{\bc}{\begin{center}}
\newcommand{\ec}{\end{center}}

\allowdisplaybreaks[4]

\title{Master Integrals for the two-loop, non-planar QCD corrections to top-quark pair production in the quark-annihilation channel}

\author[a]{Matteo Becchetti,}
\author[b,c]{Roberto Bonciani,}
\author[b,c]{Valerio Casconi,}
\author[d,e]{Andrea Ferroglia,}
\author[b,c]{Simone Lavacca,}
\author[f]{and Andreas von Manteuffel}

\affiliation[a]{Center for Cosmology, Particle Physics and Phenomenology (CP3), Universit\'e Catholique de Louvain, 1348 Louvain-La-Neuve, Belgium}
\emailAdd{matteo.becchetti@uclouvain.be}

\affiliation[b]{Dipartimento di Fisica, Sapienza - Universit\`a di Roma, Piazzale Aldo Moro 5, 00185, Rome, Italy}
\affiliation[c]{INFN Sezione di Roma, Piazzale Aldo Moro 2, 00185, Rome, Italy}
\emailAdd{roberto.bonciani@roma1.infn.it}
\emailAdd{valerio.casconi@roma1.infn.it}
\emailAdd{simone.lavacca@roma1.infn.it}

\affiliation[d]{Physics Department, New York City College of Technology, The City
	University of New York, 
	300 Jay Street, Brooklyn, NY 11201 USA}
\affiliation[e]{The Graduate School and University Center,
	The City University of New York, 365 Fifth Avenue,
	New York, NY 10016  USA}
\emailAdd{aferroglia@citytech.cuny.edu}

\affiliation[f]{Department of Physics and Astronomy, Michigan State University, 
East Lansing, MI 48824, USA}
\emailAdd{manteuffel@pa.msu.edu} 

\preprint{MSUHEP-18-021}

\abstract{We present the analytic calculation of the Master Integrals for the two-loop, non-planar topologies that enter the calculation of the amplitude for top-quark pair hadroproduction in the 
quark-annihilation channel. Using the method of differential equations, we expand the integrals in powers of the dimensional regulator $\epsilon$ and determine the expansion coefficients in terms
of generalized harmonic polylogarithms of two dimensionless variables through to weight four.}

\begin{document}

\maketitle


\section{Introduction \label{introduction}}

Theoretical predictions for top-antitop pair production at hadron colliders are known in perturbative QCD up to next-to-next-to-leading order (NNLO) \cite{Baernreuther:2012ws,Czakon:2012zr,Czakon:2012pz,Czakon:2013goa,Czakon:2014xsa,Czakon:2015owf,Czakon:2016dgf,Czakon:2016ckf}. Recently, also the NLO electroweak corrections to this process were evaluated \cite{Czakon:2017wor}.
Predictions at NNLO in QCD are available for the total cross section and for distributions that are differential with respect to quantities which depend on the momenta of the top-antitop pair, such as the pair invariant mass, the top (or antitop) transverse momentum and rapidity, etc. 

From the technical point of view, the numerical calculations carried out in  \cite{Baernreuther:2012ws,Czakon:2012zr,Czakon:2012pz,Czakon:2013goa,Czakon:2014xsa,Czakon:2015owf,Czakon:2016dgf,Czakon:2016ckf} represent a landmark in the field of the evaluation of higher-order corrections in perturbative QCD. 
One of the main technical problems that was necessary to solve in order to achieve NNLO accuracy was the evaluation of two-loop $2 \to 2$ amplitudes with massive and massless propagators. The evaluation had to be carried out for arbitrary values of the Mandelstam invariants $s$ and $t$ and of the top-quark mass $m$. The problem was solved by evaluating numerically these diagrams in a grid of points covering all of the physics phase space in the $s-t$ plane, for a fixed value of $m$. The evaluation in each single point was carried out by solving numerically differential equations satisfied by the Master Integrals (MIs) present in the problem. The numerical solution of large sets of differential equations is not only technically challenging but it also requires a significant amount of CPU time. In addition, it was necessary to evaluate analytically the boundary conditions to be used in the numerical solution of the differential equations.

In this context, an analytic calculation of the two-loop amplitudes contributing to top-quark pair production has a twofold purpose: on the one hand, it provides an independent check of the results obtained numerically; on the other hand it could provide a faster and cheaper (in terms of CPU time) way to evaluate the two-loop corrections needed in order to obtain phenomenological predictions for this process.

A complete analytic computation of the top-pair production cross section to NNLO in QCD is not yet available, although many of the necessary elements were evaluated in the recent past. In particular, the matrix elements for the one-loop $2\to 3$ process are known  \cite{Dittmaier:2007wz,Bevilacqua:2010ve,Bevilacqua:2011aa,Melnikov:2010iu}. 
Furthermore, progress was also made in the determination of infra-red (IR) subtraction terms which are  
needed to regularize IR divergences in collinear and soft regions of the phase space during the integration \cite{Abelof:2014fza,Abelof:2014jna,Abelof:2015lna,Bonciani:2015sha}. Very recently the computation of the NNLO IR subtraction terms was completed in the $Q_T$ subtraction formalism \cite{Angeles-Martinez:2018mqh,Catani:2019iny}.
Finally, the one-loop squared matrix elements were calculated in 
\cite{Korner:2008bn,Kniehl:2008fd,Anastasiou:2008vd}. Analytic results for the interference between two-loop 
$2\to 2$ diagrams and  tree-level amplitudes are available only in part.

Two-loop contributions to the $t\bar{t}$ production process in hadronic collisions are
required for two partonic channels: $q\bar{q} \to t\bar{t}$ (quark-annihilation channel) and $gg \to t\bar{t}$ (gluon fusion channel). The interference of the two-loop 
amplitude in the quark-annihilation channel with the corresponding tree-level amplitude can be expressed in terms of 
ten gauge independent functions. Each one of these functions is proportional to a different color coefficient. In the rest of this work we refer to  these functions as ``color factors''. The color structure in the  gluon-fusion channel is more complicated, and it can be expressed in
terms of sixteen color factors. 

All of the ten color factors in the $q\bar{q}$ channel are known numerically
\cite{Czakon:2008zk} and their infrared poles are known analytically \cite{Ferroglia:2009ep,Ferroglia:2009ii}.
For eight out of the ten color factors a complete analytic expression, written  in terms of generalized harmonic polylogarithms (GPLs)
\cite{Goncharov:polylog,Goncharov2007,Remiddi:1999ew,Vollinga:2004sn}, was found in 
\cite{Bonciani:2008az,Bonciani:2009nb}. 
The remaining two color factors in the quark-annihilation channel are not known analytically to date.

All of the sixteen color factors appearing in the two-loop corrections in the  gluon-fusion channel are known numerically \cite{Baernreuther:2013caa} and the analytic expression of all the infrared poles was evaluated in \cite{Ferroglia:2009ep,Ferroglia:2009ii}. In addition, a complete analytic expression (again written in terms of GPLs) is known for ten out of the sixteen color factors in the gluon fusion channel \cite{Bonciani:2010mn,vonManteuffel:2013uoa,Bonciani:2013ywa}. The remaining six color factors in this partonic channel are known to involve elliptic integrals. Very recently, the MIs for planar topologies that involve a closed heavy fermionic 
loop were studied in  \cite{Adams:2018bsn,Adams:2018kez}. These MIs contribute to one of the six gluon-channel color factors that are not known analytically.

In this paper we focus on the analytic calculation of the MIs that are needed to complete the evaluation of the two color factors in the quark-annihilation channel which are not yet known analytically. Part of the MIs needed for this task are known from previous works \cite{Bonciani:2003te,Bonciani:2003hc,Bonciani:2008wf,Bonciani:2008az,Bonciani:2009nb,Bonciani:2010mn,vonManteuffel:2013uoa,Bonciani:2013ywa} (see also the {\tt Loopedia} database~\cite{Bogner:2017xhp}).
In particular, the first analytical evaluation of a crossed double box with a massive propagator was presented in \cite{vonManteuffel:2013uoa} in terms of GPLs. More recently, within the context of a project that requires the analytic evaluation of the NNLO QED corrections to electron-muon scattering \cite{Calame:2015fva,Abbiendi:2016xup}, a planar \cite{Mastrolia:2017pfy} and a crossed \cite{DiVita:2018nnh} topology, which also enter top-pair production in the $q\bar{q}$ channel, were evaluated analytically using GPLs. 
In the present work, we provide results for the MIs belonging to the last missing crossed topology
and we carry out an independent calculation of the MIs of the topology evaluated in \cite{DiVita:2018nnh}. These results  will allow one to complete the analytic calculation of the two-loop corrections to top-quark pair production in the $q\bar{q}$ channel.

The evaluation of the MIs discussed in this work is carried out by following a by now standard technique based on two steps.
First, one observes that the dimensionally regularized scalar integrals which appear in the interference of two-loop and tree-level diagrams can all be written in terms of a reduced set of scalar integrals which are identified as the MIs for the problem under study. The two topologies considered in this work involve  52 and 44 MIs, respectively.
The reduction to MIs is carried out by means of the computer programs\footnote{Other public programs for the reduction to the MIs can be found in \cite{Anastasiou:2004vj,Lee:2012cn,Lee:2013mka,Maierhoefer:2017hyi}.}
\texttt{FIRE} \cite{Smirnov:2008iw,Smirnov:2013dia,Smirnov:2014hma} and
\texttt{Reduze 2} \cite{Studerus:2009ye,vonManteuffel:2012np}, that implement integration-by-parts identities
\cite{Tkachov:1981wb,Chetyrkin:1981qh,Laporta:2001dd} and Lorentz-invariance identities \cite{Gehrmann:1999as}.
Subsequently, the MIs are computed by employing the differential equations method \cite{Kotikov:1990kg,Remiddi:1997ny,Gehrmann:1999as,Argeri:2007up,Henn:2014qga}. The system of differential equations is cast in canonical form \cite{Henn:2013pwa} (see also \cite{Argeri:2014qva,DiVita:2014pza,Henn:2013nsa,Gehrmann:2014bfa,vonManteuffel:2014mva,Lee:2014ioa,Adams:2017tga,Schneider:2016szq,Meyer:2016slj,Georgoudis:2016wff,Gituliar:2017vzm}). The solution is expressed
in terms of Chen's iterated integrals, which can be expanded as a series in the dimensional regularization parameter, and each order of the expansion is represented in terms of GPLs.

The paper is structured as follows. In Section \ref{notations}, we introduce our notation and we define the topologies that are considered in this work. In Section \ref{Sec3}, we briefly review the method of differential equations. In Section \ref{MI}, we present the canonical form we used for the evaluation of the solution of the system of differential equations.
In Section \ref{sec:param}, we describe a reparametrization which rationalizes our differential equations. In Section \ref{sec:integration}, we discuss the integration of the differential equations in terms of GPLs and present the structure of our results. In Section \ref{numericalchecks}, we discuss numerical checks which were carried out in order to validate the analytic expression of the MIs. We emphasize that, in addition to the checks discussed in Section~\ref{numericalchecks}, our results have been successfully compared against the expressions of a different set of master integrals, independently obtained by S.~Di~Vita, T.~Gehrmann, S.~Laporta, P.~Mastrolia, A.~Primo, and  U.~Schubert \cite{PD}, which were published on the {\tt arXiv} simultaneously to the present manuscript. 
Finally, Section \ref{conclusions} contains our conclusions. The definition of the various MIs in terms of momentum integrals over a set of propagators can be found in Appendix~\ref{pre-can}.
Numerical results in a specific phase-space point for the seven denominator MIs evaluated analytically in this paper are collected in Appendix~\ref{num-res}.

Our full analytical results are provided in ancillary files included in the {\tt arXiv} submission
of this paper.

\section{Notations \label{notations}}

In this paper we consider the process $q\bar{q} \to t\bar{t}$, where $q$ and $\bar{q}$ are massless quarks and $t$ and $\bar{t}$ are massive (top) quarks. The incoming partons have momenta $p_1$ and $p_2$, while the final state partons have momenta $p_3$ and $p_4$. 
All particles are on their mass-shell, namely $p_{1}^{2}\,=\,p_{2}^{2}=0$, and $p_{3}^{2}=p_{4}^{2}=m^2$, where $m$ is top-quark mass.

The kinematics of the process can be described in terms of the three Mandelstam invariants
\begin{equation}
s=(p_{1}+p_{2})^2,\quad t=(p_{1}-p_{3})^2, \quad u = (p_1-p_4)^2 \, ,
\end{equation}
which satisfy the relation $s+t+u=2m^2$. The physical region is defined by 
\be
s > 4m^2 \, , \qquad t = m^2 - \frac{1}{2} \left( s-\sqrt{s(s-4m^2)} \cos{\theta} \right) \, ,
\ee
where $\theta$ is the scattering angle of top quark with respect to the direction of the incoming $q$ quark in the partonic center of mass frame.

Figure \ref{fig1} shows the two  seven-denominator two-loop topologies that we consider in 
this paper; they are indicated with the capital letters $A$ and $B$.
The scalar integrals belonging to Topology $A$ are defined as
\begin{equation}
\intd\frac{D_{4}^{-a_{4}}\,D_{6}^{-a_{6}}}{D_{1}^{a_{1}}\,D_{2}^{a_{2}}\,D_{3}^{a_{3}}\,D_{5}^{a_{5}}\,D_{7}^{a_{7}}\,D_{8}^{a_{8}}\,D_{9}^{a_{9}}} \, ,
\end{equation}
while the scalar integrals belonging to Topology $B$ are defined as
\begin{equation}
\intd\frac{D_{5}^{-b_{5}}\,D_{6}^{-b_{6}}}{D_{1}^{b_{1}}\,D_{2}^{b_{2}}\,D_{3}^{b_{3}}\,D_{4}^{b_{4}}\,D_{7}^{b_{7}}\,D_{8}^{b_{8}}\,D_{9}^{b_{9}}} \, .
\end{equation}
The labels $a_i$ and $b_i$, with $i=1,...,9$, are integer numbers where $a_4$, $a_6$, $b_5$, $b_6$ $\leq 0$. The $D_i$, $i=1,...,9$, are the denominators and numerators involved and $d$ is the dimension of the space-time.
The normalization of the integrals is such that
\be
{\mathcal D}^dk_i = \frac{d^d k_i}{i \pi^{\frac{d}{2}}} e^{\epsilon\gamma_E} \left( \frac{m^2}{\mu^2} \right) ^{\epsilon} \, ,
\label{eq:intmeas}
\ee
where $\epsilon = (4-d)/2$, $\gamma_E=0.5772..$ is the Euler-Mascheroni constant
and $\mu$ is the 't Hooft scale.

The nine-propagator integral family that we use for the reduction of both Topology $A$ and 
Topology $B$ is
\begin{eqnarray}
D_{\,i}&=&\lbrace -k_{1}^{2},\,-k_{2}^{2},\,-\left(p_{1}+k_{1}\right)^{2},\,-\left(p_{1}+k_{1}+k_{2} \right)^{2},\,-\left( k_{1}+p_{1}+p_{2}\right)^{2},\,-\left( k_{2}+p_{1}+p_{2}\right)^{2},\nn\\
&&-\left( k_{1}+k_{2}+p_{1}+p_{2}\right)^{2},\,m^2-\left(k_{1}+k_{2}+p_{3}\right)^{2},\,m^2-\left(k_{2}+p_{3}\right)^{2}
\rbrace \, .
\label{eq:dens}
\end{eqnarray}
%
\bfig
\bc
\[ 
\vcenter{
\hbox{
  \begin{picture}(0,0)(0,0)
\SetScale{1.0}
  \SetWidth{1}
\Line(-25,30)(-40,30)
\Line(-25,-30)(-40,-30)
\Line(-25,30)(-25,-30)
\Line(-25,30)(25,0)
\Line(0,-30)(25,30)
\Line(-25,-30)(25,-30)
%
%
  \SetWidth{4}
\Line(25,-30)(25,30)
\Line(25,30)(40,30)
\Line(25,-30)(40,-30)
\Text(0,-45)[c]{\footnotesize{(A)}}
\end{picture}}
}
\hspace*{5cm}
\vcenter{
\hbox{
  \begin{picture}(0,0)(0,0)
\SetScale{1.0}
  \SetWidth{1}
\Line(-25,30)(-40,30)
\Line(-25,-30)(-40,-30)
\Line(-25,30)(-25,-30)
\Line(-25,30)(25,0)
\Line(-25,0)(25,30)
\Line(-25,-30)(25,-30)
%
%
  \SetWidth{4}
\Line(25,-30)(25,30)
\Line(25,30)(40,30)
\Line(25,-30)(40,-30)
%
%
%
%
\Text(0,-45)[c]{\footnotesize{(B)}}
%
\end{picture}}
}
\]
\vspace*{7mm}
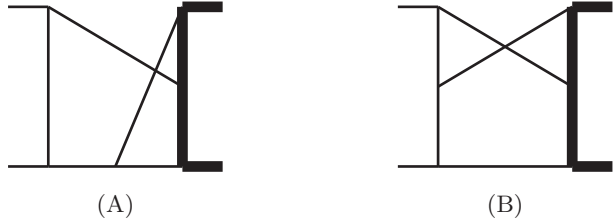
\caption{Seven-denominator topologies. Thin lines represent massless external particles and internal propagators, while thick lines represent massive external particles and internal propagators.
\label{fig1} }
\ec
\efig
%
Topology $A$ has 52 MIs while Topology $B$ has 44 MIs. Since some MIs are common to both topologies,
the total number of independent MIs is 70. Some of the MIs were already known in the literature
\cite{Bonciani:2003te,Bonciani:2003hc,Bonciani:2008wf,Bonciani:2008az,Bonciani:2009nb,Bonciani:2010mn,vonManteuffel:2013uoa,Bonciani:2013ywa,DiVita:2018nnh}. Many MIs, including many seven denominator four-point functions, are evaluated here for the first time.

\section{The Differential Equations Method}
\label{Sec3}

The analytic computation of the MIs is carried out by employing the differential equations method 
\cite{Kotikov:1990kg,Remiddi:1997ny,Gehrmann:2002zr,Argeri:2007up,Henn:2014qga}. 
The MIs can be thought of as components of a vector $\vec{f}$ where each component depends on a vector $\vec{x}$ of dimensionless parameters and  on the dimensional regulator $\epsilon$.
The dimensionless parameters in $\vec{x}$ are functions of the kinematic invariants of the problem. In the case under study, the vector $\vec{x}$ has two components; the specific choice of these components is discussed in Section~\ref{sec:param}. The MIs $\vec{f}(\vec{x},\epsilon)$ satisfy a system of first-order partial linear differential 
equations with respect to the kinematic invariants $\vec{x}$:
\begin{equation}
\label{EqDiffFundamental}
\partial_i \vec{f}(\vec{x},\epsilon)=A_i(\vec{x},\epsilon)\vec{f}(\vec{x},\epsilon) \, ,
\end{equation}
where $\partial_{i}=\partial/\partial x_{i}$, $A_i(\vec{x},\epsilon)$ is a set of matrices  associated with the system of differential equations.
These matrices have dimensions $N \times N$, where $N$ is the number of MIs in the vector $\vec{f}$.
In general, the elements of $A_i$ also depend on the kinematic invariants, $\vec{x}$, and on the dimensional regulator $\epsilon$. The matrices $A_i(\vec{x},\epsilon)$ satisfy the integrability conditions
\begin{equation}
\partial_{i}A_{j}-\partial_{j}A_{i} + \left[A_{i},A_{j}\right]=0 \, ,
\end{equation}
where  $\left[ A_{i}, A_{j} \right]\equiv A_{i}A_{j}-A_{j}A_{i}$ is the usual matrix commutator.
For a given choice of MIs, the matrices $A_i$ can be computed using integration-by-parts identities.

We solve the system in \eqref{EqDiffFundamental} by employing the Canonical Basis approach \cite{Henn:2013pwa, Henn:2014qga}, which consists in finding a basis for the MIs in which the system of differential equations has the specific form 
\be
\label{CanSys}
d\vec{f}(\vec{x},\epsilon) = \epsilon\, d\tilde{A}(\vec{x})\,\vec{f}(\vec{x},\epsilon) \, .
\ee
In (\ref{CanSys}), $d\tilde{A}(\vec{x})$ is a logarithmic differential one-form.
Several methods that allow one to find a Canonical Basis for a given topology have been proposed \cite{Henn:2013pwa, Henn:2014qga, Argeri:2014qva, Lee:2014ioa, Georgoudis:2016wff,Meyer:2017joq}. In this work, we find the Canonical Basis by employing  the semi-algorithmic approach described in \cite{Gehrmann:2014bfa, Becchetti:2017abb}.

In this basis the solution of the system of differential equations in \eqref{CanSys} is formally written in terms of Chen iterated integrals \cite{Chen:1977oja}:
\begin{equation}
\label{ChenInt}
\vec{f}(\vec{x},\epsilon)=\mathbb{P}\, \exp\left(\epsilon\int_{\gamma}d\tilde{A}(\vec{x}) \right)\vec{f}_{0}(\epsilon) \, ,
\end{equation} 
where $\mathbb{P}$ stands for the path-ordered integration, $\gamma$ is some path in the space of kinematic invariants and $\vec{f}_{0}(\epsilon)$ is a vector of boundary conditions that we found by imposing the regularity
of the MIs in particular points of the phase space or known solutions for simple integrals.

For the process under study it is possible to find a change of variables such that the matrix $\tilde{A}(\vec{x})$ is a rational function of the kinematic invariants, i.e. the entries of the one-form $d\tilde{A}(\vec{x})$ are linear combinations of terms $d\log(x_{k}-\alpha_{k})$, where $\alpha_{k}$ are algebraic functions of kinematic invariants and the arguments of the logarithms $(x_k - \alpha_k)$ determine the so called \emph{alphabet} of the process. 
In this case, once a path $\gamma=\gamma(t)$ is specified, the solution can be written order-by-order in the dimensional regularization parameter explicitly 
in terms of GPLs
\begin{equation}
 G(\alpha_{1},\dots,\alpha_{n};z)=
 \int_{0}^{z}\frac{dt}{t-\alpha_{1}} G(\alpha_{2},\dots,\alpha_{n};t) \, ,
 \end{equation}  
with
\begin{equation}
G(\alpha_{1};z)=\int_{0}^{z}\frac{dt}{t-\alpha_{1}} \quad \mbox{for} \quad \alpha_{1}\neq 0,
\qquad \mbox{and} \quad G(\vec{0}_{n};z)=\frac{\log^{n}(z)}{n!} \, ,
\end{equation}
where $\vec{0}_{n}$ indicates a list of $n$ weights, all equal to $0$.

\section{Canonical Form for the Master Integrals \label{MI}}

\bfig
\bc
\vspace*{-0.5cm}
\[ \vcenter{
\hbox{
  \begin{picture}(0,0)(0,0)
\SetScale{0.4}
  \SetWidth{1}
\CCirc(-15,15){5}{0.9}{0.9}
\CCirc(15,15){5}{0.9}{0.9}
  \SetWidth{4}
%
\CArc(-15,0)(15,0,180)
\CArc(-15,0)(15,180,360)
\CArc(15,0)(15,180,360)
%
\CArc(15,0)(15,0,180)
\Text(0,-20)[c]{\footnotesize{(${\mathcal T}^{A}_{1},{\mathcal T}^{B}_{1}$)}}
\end{picture}}
}
\hspace{1.4cm}
\vcenter{
\hbox{
  \begin{picture}(0,0)(0,0)
\SetScale{0.4}
  \SetWidth{1.0}
\DashLine(-35,0)(-20,0){3}
\DashLine(20,0)(35,0){3}
\CArc(0,0)(20,0,180)
\CArc(0,0)(20,180,360)
\CCirc(0,20){5}{0.9}{0.9}
\CCirc(34.5,23.5){5}{0.9}{0.9}
  \SetWidth{4}
\CArc(60,0)(40,150,180)
\CArc(0,34.6)(40,300,330)
\CArc(30.20,17.60)(5.28,-34,153)
\Text(-13,10)[c]{\footnotesize{$t$}}
\Text(0,-20)[c]{\footnotesize{(${\mathcal T}^{A}_9$)}}
\end{picture}}
}
\hspace{1.55cm}
\vcenter{
\hbox{
  \begin{picture}(0,0)(0,0)
\SetScale{0.4}
  \SetWidth{1.0}
\DashLine(-35,0)(-20,0){3}
\DashLine(20,0)(35,0){3}
\CCirc(0,20){5}{0.9}{0.9}
\CCirc(34.5,23.5){5}{0.9}{0.9}
\CArc(0,0)(20,180,360)
  \SetWidth{4}
\CArc(0,0)(20,0,180)
\CArc(60,0)(40,150,180)
\CArc(0,34.6)(40,300,330)
\CArc(30.20,17.60)(5.28,-34,153)
\Text(-13,10)[c]{\footnotesize{$t$}}
\Text(0,-20)[c]{\footnotesize{(${\mathcal T}^{B}_2$)}}
\end{picture}}
}
\hspace{1.55cm}
\vcenter{
\hbox{
  \begin{picture}(0,0)(0,0)
\SetScale{0.4}
  \SetWidth{1.0}
\DashLine(-35,0)(-20,0){3}
\DashLine(20,0)(35,0){3}
\CCirc(0,20){5}{0.9}{0.9}
\CCirc(0,0){5}{0.9}{0.9}
\Line(-20,0)(20,0)
  \SetWidth{4}
\CArc(0,0)(20,0,180)
\CArc(0,0)(20,180,360)
\Text(-13,10)[c]{\footnotesize{$s$}}
\Text(0,-20)[c]{\footnotesize{(${\mathcal T}^{A}_{2}$)}}
\end{picture}}
}
\hspace{1.55cm}
\vcenter{
\hbox{
  \begin{picture}(0,0)(0,0)
\SetScale{0.4}
  \SetWidth{1.0}
\DashLine(-35,0)(-20,0){3}
\DashLine(20,0)(35,0){3}
\CCirc(0,20){5}{0.9}{0.9}
\CCirc(0,-20){5}{0.9}{0.9}
\Line(-20,0)(20,0)
  \SetWidth{4}
\CArc(0,0)(20,0,180)
\CArc(0,0)(20,180,360)
\Text(-15,10)[c]{\footnotesize{$s$}}
\Text(0,-20)[c]{\footnotesize{(${\mathcal T}^{A}_3$)}}
\end{picture}}
}
\hspace{1.55cm}
\vcenter{
\hbox{
  \begin{picture}(0,0)(0,0)
\SetScale{0.4}
  \SetWidth{1.0}
\DashLine(-35,0)(-20,0){3}
\DashLine(20,0)(35,0){3}
\CCirc(0,20){5}{0.9}{0.9}
\CCirc(0,0){5}{0.9}{0.9}
\CArc(0,0)(20,0,180)
\CArc(0,0)(20,180,360)
  \SetWidth{4}
\Line(-20,0)(20,0)
\Text(-13,10)[c]{\footnotesize{$p_3^2$}}
\Text(0,-20)[c]{\footnotesize{(${\mathcal T}^{A}_4,{\mathcal T}^{B}_5$)}}
\end{picture}}
}
\hspace{1.55cm}
\vcenter{
\hbox{
  \begin{picture}(0,0)(0,0)
\SetScale{0.4}
  \SetWidth{1.0}
\DashLine(-35,0)(-20,0){3}
\DashLine(20,0)(35,0){3}
\CCirc(0,20){5}{0.9}{0.9}
\CCirc(0,-20){5}{0.9}{0.9}
\CArc(0,0)(20,0,180)
\CArc(0,0)(20,180,360)
  \SetWidth{4}
\Line(-20,0)(20,0)
\Text(-15,10)[c]{\footnotesize{$u$}}
\Text(0,-20)[c]{\footnotesize{(${\mathcal T}^{A}_{6},{\mathcal T}^{B}_{4}$)}}
\end{picture}}
}
\hspace{1.55cm}
\vcenter{
\hbox{
  \begin{picture}(0,0)(0,0)
\SetScale{0.4}
  \SetWidth{1.0}
\DashLine(-35,0)(-20,0){3}
\DashLine(20,0)(35,0){3}
\CCirc(0,20){5}{0.9}{0.9}
\CCirc(0,0){5}{0.9}{0.9}
\CArc(0,0)(20,0,180)
\CArc(0,0)(20,180,360)
  \SetWidth{4}
\Line(-20,0)(20,0)
\Text(-13,10)[c]{\footnotesize{$u$}}
\Text(0,-20)[c]{\footnotesize{(${\mathcal T}^{A}_5,{\mathcal T}^{B}_3$)}}
\end{picture}}
}
%
\]
\vspace*{0.4cm}
\[
\vcenter{
\hbox{
  \begin{picture}(0,0)(0,0)
\SetScale{0.4}
  \SetWidth{1.0}
\DashLine(-35,0)(-20,0){3}
\DashLine(20,0)(35,0){3}
\CCirc(0,20){5}{0.9}{0.9}
\CCirc(0,-20){5}{0.9}{0.9}
\CArc(0,0)(20,0,180)
\CArc(0,0)(20,180,360)
  \SetWidth{4}
\Line(-20,0)(20,0)
\Text(-15,10)[c]{\footnotesize{$t$}}
\Text(0,-20)[c]{\footnotesize{(${\mathcal T}^{A}_{8},{\mathcal T}^{B}_{7}$)}}
\end{picture}}
}
\hspace{1.55cm}
\vcenter{
\hbox{
  \begin{picture}(0,0)(0,0)
\SetScale{0.4}
  \SetWidth{1.0}
\DashLine(-35,0)(-20,0){3}
\DashLine(20,0)(35,0){3}
\CCirc(0,20){5}{0.9}{0.9}
\CCirc(0,0){5}{0.9}{0.9}
\CArc(0,0)(20,0,180)
\CArc(0,0)(20,180,360)
  \SetWidth{4}
\Line(-20,0)(20,0)
\Text(-13,10)[c]{\footnotesize{$t$}}
\Text(0,-20)[c]{\footnotesize{(${\mathcal T}^{A}_7,{\mathcal T}^{B}_6$)}}
\end{picture}}
}
\hspace{1.55cm}
\vcenter{
\hbox{
  \begin{picture}(0,0)(0,0)
\SetScale{0.4}
  \SetWidth{1.0}
\DashLine(-35,0)(-20,0){3}
\DashLine(20,0)(35,0){3}
\CCirc(0,20){5}{0.9}{0.9}
\CCirc(0,-20){5}{0.9}{0.9}
\Line(-20,0)(20,0)
%
%
\CArc(0,0)(20,0,180)
\CArc(0,0)(20,180,360)
\Text(-15,10)[c]{\footnotesize{$s$}}
\Text(0,-20)[c]{\footnotesize{(${\mathcal T}^{A}_{10},{\mathcal T}^{B}_{8}$)}}
\end{picture}}
}
\hspace{1.65cm}
\vcenter{
\hbox{
  \begin{picture}(0,0)(0,0)
\SetScale{0.4}
  \SetWidth{1}
\DashLine(-50,0)(-35,0){3}
\CCirc(-13,-15){5}{0.9}{0.9}
\CArc(-25,-34)(35,10,105)
\Line(-35,0)(10,30)
\Line(10,-30)(-35,0)
\SetWidth{4}
\Line(10,30)(25,30)
\Line(10,-30)(25,-30)
\Line(10,30)(10,-30)
\Text(-13,10)[c]{\footnotesize{$s$}}
\Text(0,-26)[c]{\footnotesize{(${\mathcal T}^{A}_{19},{\mathcal T}^{B}_{16}$)}}
\end{picture}}
}
\hspace{1.55cm}
\vcenter{
\hbox{
  \begin{picture}(0,0)(0,0)
\SetScale{0.4}
  \SetWidth{1}
\DashLine(-50,0)(-35,0){3}
\Line(10,-30)(25,-30)
\Line(10,-30)(-35,0)
\CCirc(10,0){5}{0.9}{0.9}
\CArc(30,0)(35,125,235)
\Line(10,30)(10,-30)
\SetWidth{4}
\Line(10,30)(25,30)
\Line(-35,0)(10,30)
\Text(-13,8)[c]{\footnotesize{$t$}}
\Text(0,-26)[c]{\footnotesize{(${\mathcal T}^{B}_{17}$)}}
\end{picture}}
}
\hspace{1.55cm}
\vcenter{
\hbox{
  \begin{picture}(0,0)(0,0)
\SetScale{0.4}
  \SetWidth{1}
\DashLine(-50,0)(-35,0){3}
\CCirc(-13,-15){5}{0.9}{0.9}
\CArc(-25,-34)(35,10,105)
\Line(10,30)(10,-30)
\SetWidth{4}
\Line(10,30)(25,30)
\Line(10,-30)(25,-30)
\Line(-35,0)(10,30)
\Line(10,-30)(-35,0)
\Text(-13,10)[c]{\footnotesize{$s$}}
\Text(0,-26)[c]{\footnotesize{(${\mathcal T}^{A}_{11}$)}}
\end{picture}}
}
\hspace{1.55cm}
\vcenter{
\hbox{
  \begin{picture}(0,0)(0,0)
\SetScale{0.4}
  \SetWidth{1}
\DashLine(-50,0)(-35,0){3}
\Line(10,-30)(25,-30)
\CCirc(-13,-15){5}{0.9}{0.9}
\CArc(-25,-34)(35,10,105)
\Line(-35,0)(10,30)
\SetWidth{4}
\Line(10,30)(25,30)
\Line(10,30)(10,-30)
\Line(10,-30)(-35,0)
\Text(-13,10)[c]{\footnotesize{$u$}}
\Text(0,-26)[c]{\footnotesize{(${\mathcal T}^{A}_{12},{\mathcal T}^{B}_{9}$)}}
\end{picture}}
}
\hspace{1.55cm}
\vcenter{
\hbox{
  \begin{picture}(0,0)(0,0)
\SetScale{0.4}
  \SetWidth{1}
\DashLine(-50,0)(-35,0){3}
\Line(10,-30)(25,-30)
\CCirc(-13,-15){5}{0.9}{0.9}
\CArc(-25,-34)(35,10,105)
\Line(-35,0)(10,30)
\SetWidth{4}
\Line(10,30)(25,30)
\Line(10,30)(10,-30)
\Line(10,-30)(-35,0)
\Text(-13,10)[c]{\footnotesize{$u$}}
\Text(-9,-13)[c]{\footnotesize{$3$}}
\Text(0,-26)[c]{\footnotesize{(${\mathcal T}^{A}_{13},{\mathcal T}^{B}_{10}$)}}
\end{picture}}
}
\]
\vspace*{0.6cm}
\[
\vcenter{
\hbox{
  \begin{picture}(0,0)(0,0)
\SetScale{0.4}
  \SetWidth{1}
\DashLine(-50,0)(-35,0){3}
\Line(10,-30)(25,-30)
\CCirc(-13,-15){5}{0.9}{0.9}
\CArc(-25,-34)(35,10,105)
\Line(-35,0)(10,30)
\SetWidth{4}
\Line(10,30)(25,30)
\Line(10,30)(10,-30)
\Line(10,-30)(-35,0)
\Text(-13,10)[c]{\footnotesize{$t$}}
\Text(0,-26)[c]{\footnotesize{(${\mathcal T}^{A}_{14},{\mathcal T}^{B}_{14}$)}}
\end{picture}}
}
\hspace{1.55cm}
\vcenter{
\hbox{
  \begin{picture}(0,0)(0,0)
\SetScale{0.4}
  \SetWidth{1}
\DashLine(-50,0)(-35,0){3}
\Line(10,-30)(25,-30)
\CCirc(-13,-15){5}{0.9}{0.9}
\CArc(-25,-34)(35,10,105)
\Line(-35,0)(10,30)
\SetWidth{4}
\Line(10,30)(25,30)
\Line(10,30)(10,-30)
\Line(10,-30)(-35,0)
\Text(-13,10)[c]{\footnotesize{$t$}}
\Text(-9,-13)[c]{\footnotesize{$3$}}
\Text(0,-26)[c]{\footnotesize{(${\mathcal T}^{A}_{15},{\mathcal T}^{B}_{15}$)}}
\end{picture}}
}
\hspace{1.55cm}
\vcenter{
\hbox{
  \begin{picture}(0,0)(0,0)
\SetScale{0.4}
  \SetWidth{1}
\DashLine(-50,0)(-35,0){3}
\Line(-35,0)(10,30)
\Line(10,-30)(-35,0)
\CArc(30,0)(35,125,235)
\CCirc(10,0){5}{0.9}{0.9}
\SetWidth{4}
\Line(10,30)(25,30)
\Line(10,-30)(25,-30)
\Line(10,30)(10,-30)
\Text(-13,8)[c]{\footnotesize{$s$}}
\Text(0,-26)[c]{\footnotesize{(${\mathcal T}^{A}_{16}$)}}
\end{picture}}
}
\hspace{1.55cm}
\vcenter{
\hbox{
  \begin{picture}(0,0)(0,0)
\SetScale{0.4}
  \SetWidth{1}
\DashLine(-50,0)(-35,0){3}
\Line(-35,0)(10,30)
\Line(10,-30)(-35,0)
\CArc(30,0)(35,125,235)
\CCirc(10,0){5}{0.9}{0.9}
\SetWidth{4}
\Line(10,30)(25,30)
\Line(10,-30)(25,-30)
\Line(10,30)(10,-30)
\Text(-13,8)[c]{\footnotesize{$s$}}
\Text(10,0)[c]{\footnotesize{$3$}}
\Text(0,-26)[c]{\footnotesize{(${\mathcal T}^{A}_{17}$)}}
\end{picture}}
}
\hspace{1.55cm}
\vcenter{
\hbox{
  \begin{picture}(0,0)(0,0)
\SetScale{0.4}
  \SetWidth{1}
\DashLine(-50,0)(-35,0){3}
\Line(-35,0)(10,30)
\Line(10,-30)(-35,0)
\CArc(30,0)(35,125,235)
\CCirc(10,0){5}{0.9}{0.9}
\CCirc(-5,0){5}{0.9}{0.9}
\SetWidth{4}
\Line(10,30)(25,30)
\Line(10,-30)(25,-30)
\Line(10,30)(10,-30)
\Text(-13,8)[c]{\footnotesize{$s$}}
\Text(0,-26)[c]{\footnotesize{(${\mathcal T}^{A}_{18}$)}}
\end{picture}}
}
\hspace{1.55cm}
\vcenter{
\hbox{
  \begin{picture}(0,0)(0,0)
\SetScale{0.4}
  \SetWidth{1}
\DashLine(-50,0)(-35,0){3}
\Line(-35,0)(10,30)
\Line(10,-30)(25,-30)
\CCirc(10,0){5}{0.9}{0.9}
\CArc(30,0)(35,125,235)
\SetWidth{4}
\Line(10,30)(25,30)
\Line(10,30)(10,-30)
\Line(10,-30)(-35,0)
\Text(-13,8)[c]{\footnotesize{$t$}}
\Text(0,-26)[c]{\footnotesize{(${\mathcal T}^{B}_{11}$)}}
\end{picture}}
}
\hspace{1.55cm}
\vcenter{
\hbox{
  \begin{picture}(0,0)(0,0)
\SetScale{0.4}
  \SetWidth{1}
\DashLine(-50,0)(-35,0){3}
\Line(-35,0)(10,30)
\Line(10,-30)(25,-30)
\CCirc(10,0){5}{0.9}{0.9}
\CArc(30,0)(35,125,235)
\SetWidth{4}
\Line(10,30)(25,30)
\Line(10,30)(10,-30)
\Line(10,-30)(-35,0)
\Text(-13,8)[c]{\footnotesize{$t$}}
\Text(10,0)[c]{\footnotesize{$3$}}
\Text(0,-26)[c]{\footnotesize{(${\mathcal T}^{B}_{12}$)}}
\end{picture}}
}
\hspace{1.55cm}
\vcenter{
\hbox{
  \begin{picture}(0,0)(0,0)
\SetScale{0.4}
  \SetWidth{1}
\DashLine(-50,0)(-35,0){3}
\Line(-35,0)(10,30)
\Line(10,-30)(25,-30)
\CCirc(10,0){5}{0.9}{0.9}
\CCirc(-16,-12){5}{0.9}{0.9}
\CArc(30,0)(35,125,235)
\SetWidth{4}
\Line(10,30)(25,30)
\Line(10,30)(10,-30)
\Line(10,-30)(-35,0)
\Text(-13,8)[c]{\footnotesize{$t$}}
\Text(0,-26)[c]{\footnotesize{(${\mathcal T}^{B}_{13}$)}}
\end{picture}}
}
\]
\vspace*{0.6cm}
\[
\vcenter{
\hbox{
  \begin{picture}(0,0)(0,0)
\SetScale{0.4}
  \SetWidth{1}
\Line(-25,30)(-40,30)
\Line(-25,-30)(-40,-30)
\Line(-25,-30)(25,-30)
\Line(-25,30)(-25,-30)
\Line(-25,-30)(25,30)
\Line(25,-30)(-25,30)
%
  \SetWidth{4}
\Line(25,-30)(40,-30)
\Line(25,30)(40,30)
\Line(25,-30)(25,30)
%
%
%
\Text(0,-26)[c]{\footnotesize{(${\mathcal T}^{A}_{36},{\mathcal T}^{B}_{33}$)}}
%
\end{picture}}
}
\hspace{1.55cm}
\vcenter{
\hbox{
  \begin{picture}(0,0)(0,0)
\SetScale{0.4}
  \SetWidth{1}
\Line(-25,30)(-40,30)
\Line(-25,-30)(-40,-30)
\Line(-25,-30)(25,-30)
\Line(-25,30)(-25,-30)
\Line(-25,-30)(25,30)
\Line(25,-30)(-25,30)
\CCirc(25,0){5}{0.9}{0.9}
  \SetWidth{4}
\Line(25,-30)(40,-30)
\Line(25,30)(40,30)
\Line(25,-30)(25,30)
%
%
%
\Text(0,-26)[c]{\footnotesize{(${\mathcal T}^{A}_{37},{\mathcal T}^{B}_{34}$)}}
%
\end{picture}}
}
\hspace{1.55cm}
\vcenter{
\hbox{
  \begin{picture}(0,0)(0,0)
\SetScale{0.4}
  \SetWidth{1}
\Line(-25,30)(-40,30)
\Line(-25,-30)(-40,-30)
\Line(-25,-30)(25,-30)
\Line(-25,30)(-25,-30)
\Line(-25,-30)(25,30)
\Line(25,30)(-25,30)
%
  \SetWidth{4}
\Line(25,-30)(40,-30)
\Line(25,30)(40,30)
\Line(25,-30)(25,30)
%
%
%
\Text(0,-26)[c]{\footnotesize{(${\mathcal T}^{A}_{38},{\mathcal T}^{B}_{30}$)}}
%
\end{picture}}
}
\hspace{1.55cm}
\vcenter{
\hbox{
  \begin{picture}(0,0)(0,0)
\SetScale{0.4}
  \SetWidth{1}
\Line(-25,30)(-40,30)
\Line(-25,-30)(-40,-30)
\Line(-25,-30)(25,-30)
\Line(-25,30)(-25,-30)
\Line(-25,-30)(25,30)
\Line(25,30)(-25,30)
\CCirc(25,0){5}{0.9}{0.9}
  \SetWidth{4}
\Line(25,-30)(40,-30)
\Line(25,30)(40,30)
\Line(25,-30)(25,30)
%
%
%
\Text(0,-26)[c]{\footnotesize{(${\mathcal T}^{A}_{39},{\mathcal T}^{B}_{31}$)}}
%
\end{picture}}
}
\hspace{1.55cm}
\vcenter{
\hbox{
  \begin{picture}(0,0)(0,0)
\SetScale{0.4}
  \SetWidth{1}
\Line(-25,30)(-40,30)
\Line(-25,-30)(-40,-30)
\Line(-25,-30)(25,-30)
\Line(-25,30)(25,30)
\Line(-25,-30)(25,30)
\Line(25,-30)(-25,30)
%
  \SetWidth{4}
\Line(25,-30)(40,-30)
\Line(25,30)(40,30)
\Line(25,-30)(25,30)
\Text(0,-26)[c]{\footnotesize{(${\mathcal T}^{B}_{24}$)}}
\end{picture}}
}
\hspace{1.55cm}
\vcenter{
\hbox{
  \begin{picture}(0,0)(0,0)
\SetScale{0.4}
  \SetWidth{1}
\Line(-25,30)(-40,30)
\Line(-25,-30)(-40,-30)
\Line(-25,-30)(25,-30)
\Line(-25,30)(25,30)
\Line(-25,-30)(25,30)
\Line(25,-30)(-25,30)
\CCirc(25,0){5}{0.9}{0.9}
  \SetWidth{4}
\Line(25,-30)(40,-30)
\Line(25,30)(40,30)
\Line(25,-30)(25,30)
\Text(0,-26)[c]{\footnotesize{(${\mathcal T}^{B}_{25}$)}}
\end{picture}}
}
\hspace{1.55cm}
\vcenter{
\hbox{
  \begin{picture}(0,0)(0,0)
\SetScale{0.4}
  \SetWidth{1}
\Line(-25,-30)(-40,-30)
\Line(-25,30)(-40,30)
\Line(-25,-30)(25,-30)
\Line(-25,30)(-25,-30)
%
\Line(25,-30)(25,30)
  \SetWidth{4}
\Line(25,-30)(40,-30)
\Line(25,30)(40,30)
\Line(-25,30)(25,-30)
\Line(25,30)(-25,30)
%
%
%
\Text(0,-26)[c]{\footnotesize{(${\mathcal T}^{A}_{27}$)}}
%
\end{picture}}
}
\hspace{1.55cm}
\vcenter{
\hbox{
  \begin{picture}(0,0)(0,0)
\SetScale{0.4}
  \SetWidth{1}
\Line(-25,-30)(-40,-30)
\Line(-25,30)(-40,30)
\Line(-25,-30)(25,-30)
\Line(-25,30)(-25,-30)
\CCirc(0,0){5}{0.9}{0.9}
\Line(25,-30)(25,30)
  \SetWidth{4}
\Line(25,-30)(40,-30)
\Line(25,30)(40,30)
\Line(-25,30)(25,-30)
\Line(25,30)(-25,30)
%
%
%
\Text(0,-26)[c]{\footnotesize{(${\mathcal T}^{A}_{28}$)}}
%
\end{picture}}
}
\]
\vspace*{0.7cm}
\[ 
\vcenter{
\hbox{
  \begin{picture}(0,0)(0,0)
\SetScale{0.4}
  \SetWidth{1}
\Line(-25,-30)(-40,-30)
\Line(-25,30)(-40,30)
\Line(-25,-30)(25,-30)
\Line(-25,30)(-25,-30)
\CCirc(0,30){5}{0.9}{0.9}
\Line(25,-30)(25,30)
  \SetWidth{4}
\Line(25,-30)(40,-30)
\Line(25,30)(40,30)
\Line(-25,30)(25,-30)
\Line(25,30)(-25,30)
%
%
%
\Text(0,-26)[c]{\footnotesize{(${\mathcal T}^{A}_{29}$)}}
%
\end{picture}}
}
\hspace{1.55cm}
\vcenter{
\hbox{
  \begin{picture}(0,0)(0,0)
\SetScale{0.4}
  \SetWidth{1}
\Line(-25,-30)(-40,-30)
\Line(-25,30)(-40,30)
\Line(-25,-30)(25,30)
\Line(-25,30)(-25,-30)
%
\Line(25,-30)(25,30)
  \SetWidth{4}
\Line(25,-30)(40,-30)
\Line(25,30)(40,30)
\Line(-25,30)(25,-30)
\Line(25,30)(-25,30)
%
%
%
\Text(0,-26)[c]{\footnotesize{(${\mathcal T}^{A}_{20}$)}}
%
\end{picture}}
}
\hspace{1.55cm}
\vcenter{
\hbox{
  \begin{picture}(0,0)(0,0)
\SetScale{0.4}
  \SetWidth{1}
\Line(-25,-30)(-40,-30)
\Line(-25,30)(-40,30)
\Line(-25,-30)(25,30)
\Line(-25,30)(-25,-30)
\CCirc(0,30){5}{0.9}{0.9}
\Line(25,-30)(25,30)
  \SetWidth{4}
\Line(25,-30)(40,-30)
\Line(25,30)(40,30)
\Line(-25,30)(25,-30)
\Line(25,30)(-25,30)
%
%
%
\Text(0,-26)[c]{\footnotesize{(${\mathcal T}^{A}_{21}$)}}
\end{picture}}
}
\hspace{1.55cm}
\vcenter{
\hbox{
  \begin{picture}(0,0)(0,0)
\SetScale{0.4}
  \SetWidth{1}
\Line(-25,-30)(-40,-30)
\Line(-25,30)(-40,30)
\Line(-25,-30)(25,30)
\Line(-25,30)(-25,-30)
\CCirc(-10,10){5}{0.9}{0.9}
\Line(25,-30)(25,30)
  \SetWidth{4}
\Line(25,-30)(40,-30)
\Line(25,30)(40,30)
\Line(-25,30)(25,-30)
\Line(25,30)(-25,30)
%
%
%
\Text(0,-26)[c]{\footnotesize{(${\mathcal T}^{A}_{22}$)}}
%
\end{picture}}
}
\hspace{1.55cm}
\vcenter{
\hbox{
  \begin{picture}(0,0)(0,0)
\SetScale{0.4}
  \SetWidth{1}
\Line(-25,30)(-40,30)
\Line(-25,-30)(-40,-30)
\Line(-25,30)(-25,-30)
\Line(-25,-30)(25,30)
\Line(25,-30)(-25,30)
%
\CCirc(25,0){5}{0.9}{0.9}
\CArc(5,0)(35,305,415)
  \SetWidth{4}
\Line(25,-30)(40,-30)
\Line(25,30)(40,30)
\Line(25,-30)(25,30)
\Text(0,-26)[c]{\footnotesize{(${\mathcal T}^{A}_{32}$)}}
\end{picture}}
}
\hspace{1.55cm}
\vcenter{
\hbox{
  \begin{picture}(0,0)(0,0)
\SetScale{0.4}
  \SetWidth{1}
\Line(-25,30)(-40,30)
\Line(-25,-30)(-40,-30)
\Line(-25,30)(-25,-30)
\Line(-25,-30)(25,30)
\Line(25,-30)(-25,30)
%
\CCirc(40,0){5}{0.9}{0.9}
\CArc(5,0)(35,305,415)
  \SetWidth{4}
\Line(25,-30)(40,-30)
\Line(25,30)(40,30)
\Line(25,-30)(25,30)
\Text(0,-26)[c]{\footnotesize{(${\mathcal T}^{A}_{33}$)}}
\end{picture}}
}
\hspace{1.55cm}
\vcenter{
\hbox{
  \begin{picture}(0,0)(0,0)
\SetScale{0.4}
  \SetWidth{1}
\Line(-25,30)(-40,30)
\Line(-25,-30)(-40,-30)
\Line(-25,30)(-25,-30)
\Line(-25,-30)(25,-30)
\Line(25,30)(-25,30)
%
\CCirc(0,18){5}{0.9}{0.9}
\CArc(0,50)(33,218,322)
  \SetWidth{4}
\Line(25,-30)(40,-30)
\Line(25,30)(40,30)
\Line(25,-30)(25,30)
\Text(0,-26)[c]{\footnotesize{(${\mathcal T}^{B}_{32}$)}}
\end{picture}}
}
\hspace{1.55cm}
\vcenter{
\hbox{
  \begin{picture}(0,0)(0,0)
\SetScale{0.4}
  \SetWidth{1.0}
\DashLine(-50,0)(-35,0){3}
\Line(10,30)(-35,0)
\Line(-35,0)(10,-30)
\Line(10,30)(10,-30)
%
  \SetWidth{4}
\Line(10,30)(25,30)
\Line(10,-30)(25,-30)
\Line(10,30)(-11,-16)
\Line(10,-30)(-11,-16)
\Text(-18,8)[c]{\footnotesize{$s$}}
\Text(0,-26)[c]{\footnotesize{(${\mathcal T}^{A}_{30}$)}}
\end{picture}}
}
\]
\vspace*{0.6cm}
\[
\vcenter{
\hbox{
  \begin{picture}(0,0)(0,0)
\SetScale{0.4}
  \SetWidth{1.0}
\DashLine(-50,0)(-35,0){3}
\Line(10,30)(-35,0)
\Line(-35,0)(10,-30)
\Line(10,30)(10,-30)
\CCirc(-5,0){5}{0.9}{0.9}
  \SetWidth{4}
\Line(10,30)(25,30)
\Line(10,-30)(25,-30)
\Line(10,30)(-11,-16)
\Line(10,-30)(-11,-16)
\Text(-18,8)[c]{\footnotesize{$s$}}
\Text(0,-26)[c]{\footnotesize{(${\mathcal T}^{A}_{31}$)}}
\end{picture}}
}
\hspace{1.55cm}
\vcenter{
\hbox{
  \begin{picture}(0,0)(0,0)
\SetScale{0.4}
  \SetWidth{1.0}
\DashLine(-50,0)(-35,0){3}
\Line(10,30)(-35,0)
\Line(-35,0)(10,-30)
\Line(-35,0)(10,0)
%
  \SetWidth{4}
\Line(10,30)(25,30)
\Line(10,-30)(25,-30)
\Line(10,30)(10,-30)
\Text(-18,8)[c]{\footnotesize{$s$}}
\Text(0,-26)[c]{\footnotesize{(${\mathcal T}^{A}_{34},{\mathcal T}^{B}_{27}$)}}
\end{picture}}
}
\hspace{1.55cm}
\vcenter{
\hbox{
  \begin{picture}(0,0)(0,0)
\SetScale{0.4}
  \SetWidth{1.0}
\DashLine(-50,0)(-35,0){3}
\Line(10,30)(-35,0)
\Line(-35,0)(10,-30)
\Line(-35,0)(10,0)
\CCirc(-10,-15){5}{0.9}{0.9}
  \SetWidth{4}
\Line(10,30)(25,30)
\Line(10,-30)(25,-30)
\Line(10,30)(10,-30)
\Text(-18,8)[c]{\footnotesize{$s$}}
\Text(0,-26)[c]{\footnotesize{(${\mathcal T}^{A}_{35},{\mathcal T}^{B}_{26}$)}}
\end{picture}}
}
\hspace{1.55cm}
\vcenter{
\hbox{
  \begin{picture}(0,0)(0,0)
\SetScale{0.4}
  \SetWidth{1.0}
\DashLine(-50,0)(-35,0){3}
\Line(-35,0)(10,-30)
\Line(10,30)(10,-30)
\Line(10,-30)(25,-30)
\Line(10,-30)(-11,16)
  \SetWidth{4}
\Line(10,30)(25,30)
\Line(10,30)(-35,0)
\Text(-18,8)[c]{\footnotesize{$t$}}
\Text(0,-26)[c]{\footnotesize{(${\mathcal T}^{B}_{28}$)}}
\end{picture}}
}
\hspace{1.55cm}
\vcenter{
\hbox{
  \begin{picture}(0,0)(0,0)
\SetScale{0.4}
  \SetWidth{1.0}
\DashLine(-50,0)(-35,0){3}
\Line(-35,0)(10,-30)
\Line(10,30)(10,-30)
\Line(10,-30)(25,-30)
\CCirc(-22,9){5}{0.9}{0.9}
\Line(10,-30)(-11,16)
  \SetWidth{4}
\Line(10,30)(25,30)
\Line(10,30)(-35,0)
\Text(-18,8)[c]{\footnotesize{$t$}}
\Text(0,-26)[c]{\footnotesize{(${\mathcal T}^{B}_{29}$)}}
\end{picture}}
}
\hspace{1.55cm}
\vcenter{
\hbox{
  \begin{picture}(0,0)(0,0)
\SetScale{0.4}
  \SetWidth{1}
\Line(-25,-30)(-40,-30)
\Line(-25,30)(-40,30)
\Line(-25,-30)(25,-30)
\Line(-25,30)(-25,-30)
%
\Line(-25,-30)(25,30)
  \SetWidth{4}
\Line(25,-30)(40,-30)
\Line(25,30)(40,30)
\Line(-25,30)(25,-30)
\Line(25,30)(-25,30)
%
%
%
\Text(0,-26)[c]{\footnotesize{(${\mathcal T}^{A}_{23},{\mathcal T}^{B}_{20}$)}}
%
\end{picture}}
}
\hspace{1.55cm}
\vcenter{
\hbox{
  \begin{picture}(0,0)(0,0)
\SetScale{0.4}
  \SetWidth{1}
\Line(-25,-30)(-40,-30)
\Line(-25,30)(-40,30)
\Line(-25,-30)(25,-30)
\Line(-25,30)(-25,-30)
\CCirc(0,30){5}{0.9}{0.9}
\Line(-25,-30)(25,30)
  \SetWidth{4}
\Line(25,-30)(40,-30)
\Line(25,30)(40,30)
\Line(-25,30)(25,-30)
\Line(25,30)(-25,30)
%
%
%
\Text(0,-26)[c]{\footnotesize{(${\mathcal T}^{A}_{24},{\mathcal T}^{B}_{21}$)}}
%
\end{picture}}
}
\hspace{1.55cm}
\vcenter{
\hbox{
  \begin{picture}(0,0)(0,0)
\SetScale{0.4}
  \SetWidth{1}
\Line(-25,-30)(-40,-30)
\Line(-25,30)(-40,30)
\Line(-25,-30)(25,-30)
\Line(-25,30)(-25,-30)
\CCirc(0,-30){5}{0.9}{0.9}
\Line(-25,-30)(25,30)
  \SetWidth{4}
\Line(25,-30)(40,-30)
\Line(25,30)(40,30)
\Line(-25,30)(25,-30)
\Line(25,30)(-25,30)
%
%
%
\Text(0,-26)[c]{\footnotesize{(${\mathcal T}^{A}_{25},{\mathcal T}^{B}_{22}$)}}
%
\end{picture}}
}
\]
\vspace*{0.9cm}
\[
\vcenter{
\hbox{
  \begin{picture}(0,0)(0,0)
\SetScale{0.4}
  \SetWidth{1}
\Line(-25,-30)(-40,-30)
\Line(-25,30)(-40,30)
\Line(-25,-30)(25,-30)
\Line(-25,30)(-25,-30)
\CCirc(12,-15){5}{0.9}{0.9}
\Line(-25,-30)(25,30)
  \SetWidth{4}
\Line(25,-30)(40,-30)
\Line(25,30)(40,30)
\Line(-25,30)(25,-30)
\Line(25,30)(-25,30)
%
%
%
\Text(0,-26)[c]{\footnotesize{(${\mathcal T}^{A}_{26},{\mathcal T}^{B}_{23}$)}}
%
\end{picture}}
}
\hspace{1.55cm}
\vcenter{
\hbox{
  \begin{picture}(0,0)(0,0)
\SetScale{0.4}
  \SetWidth{1}
\Line(-25,30)(-40,30)
\Line(-25,-30)(-40,-30)
\Line(-25,30)(-25,-30)
\Line(-25,-30)(25,-30)
\Line(25,30)(-25,30)
%
\CCirc(25,0){5}{0.9}{0.9}
\CArc(45,0)(35,125,235)
  \SetWidth{4}
\Line(25,-30)(40,-30)
\Line(25,30)(40,30)
\Line(25,-30)(25,30)
\Text(0,-26)[c]{\footnotesize{(${\mathcal T}^{A}_{40}$)}}
\end{picture}}
}
\hspace{1.7cm}
\vcenter{
\hbox{
  \begin{picture}(0,0)(0,0)
\SetScale{0.4}
  \SetWidth{1}
\Line(-25,30)(-40,30)
\Line(-25,-30)(-40,-30)
\Line(-25,30)(-25,-30)
\Line(-25,-30)(25,-30)
\Line(25,30)(-25,30)
\CCirc(25,0){5}{0.9}{0.9}
\CArc(45,0)(35,125,235)
  \SetWidth{4}
\Line(25,-30)(40,-30)
\Line(25,30)(40,30)
\Line(25,-30)(25,30)
\Text(0,-26)[c]{\footnotesize{(${\mathcal T}^{A}_{41}$)}}
\Text(16,0)[c]{\footnotesize{$3$}}
\end{picture}}
}
\hspace{1.55cm}
\vcenter{
\hbox{
  \begin{picture}(0,0)(0,0)
\SetScale{0.4}
  \SetWidth{1}
\Line(-25,30)(-40,30)
\Line(-25,-30)(-40,-30)
\Line(-25,-30)(25,-30)
\Line(-25,-30)(25,30)
%
\CCirc(0,30){5}{0.9}{0.9}
\CArc(0,10)(33,38,142)
  \SetWidth{4}
\Line(25,-30)(40,-30)
\Line(25,30)(40,30)
\Line(25,30)(-25,30)
\Line(-25,30)(25,-30)
\Text(0,-26)[c]{\footnotesize{(${\mathcal T}^{B}_{18}$)}}
\end{picture}}
}
\hspace{1.55cm}
\vcenter{
\hbox{
  \begin{picture}(0,0)(0,0)
\SetScale{0.4}
  \SetWidth{1}
\Line(-25,30)(-40,30)
\Line(-25,-30)(-40,-30)
\Line(-25,-30)(25,-30)
\Line(-25,-30)(25,30)
%
\CCirc(0,30){5}{0.9}{0.9}
\CArc(0,10)(33,38,142)
  \SetWidth{4}
\Line(25,-30)(40,-30)
\Line(25,30)(40,30)
\Line(25,30)(-25,30)
\Line(-25,30)(25,-30)
\Text(0,-26)[c]{\footnotesize{(${\mathcal T}^{B}_{19}$)}}
\Text(0,22)[c]{\footnotesize{$3$}}
\end{picture}}
}
\hspace{1.55cm}
\vcenter{
\hbox{
  \begin{picture}(0,0)(0,0)
\SetScale{0.4}
  \SetWidth{1.0}
\DashLine(-50,0)(-35,0){3}
\Line(10,30)(-16,-11)
\Line(-35,0)(10,-30)
\Line(-35,0)(10,0)
%
  \SetWidth{4}
\Line(10,30)(25,30)
\Line(10,-30)(25,-30)
\Line(10,30)(10,-30)
\Text(-18,8)[c]{\footnotesize{$s$}}
\Text(0,-26)[c]{\footnotesize{(${\mathcal T}^{A}_{43}$)}}
\end{picture}}
}
\hspace{1.55cm}
\vcenter{
\hbox{
  \begin{picture}(0,0)(0,0)
\SetScale{0.4}
  \SetWidth{1}
\Line(-25,30)(-40,30)
\Line(-25,-30)(-40,-30)
\Line(-25,30)(-25,-30)
\Line(-25,-30)(25,30)
\Line(25,-10)(-25,30)
%
\CArc(5,0)(35,305,415)
  \SetWidth{4}
\Line(25,-30)(40,-30)
\Line(25,30)(40,30)
\Line(25,-30)(25,30)
\Text(0,-26)[c]{\footnotesize{(${\mathcal T}^{A}_{42}$)}}
\end{picture}}
}
\hspace{1.55cm}
\vcenter{
\hbox{
  \begin{picture}(0,0)(0,0)
\SetScale{0.4}
  \SetWidth{1}
\Line(-25,30)(-40,30)
\Line(-25,-30)(-40,-30)
\Line(-25,-30)(25,-30)
\Line(-25,30)(25,30)
\Line(-25,-30)(-25,30)
\Line(25,-30)(25,30)
%
  \SetWidth{4}
\Line(25,-30)(40,-30)
\Line(25,30)(40,30)
\Line(0,30)(25,30)
\Line(0,30)(25,-30)
%
%
%
\Text(0,-26)[c]{\footnotesize{(${\mathcal T}^{A}_{48}$)}}
%
\end{picture}}
}
\]
\vspace*{0.9cm}
\[
\vcenter{
\hbox{
  \begin{picture}(0,0)(0,0)
\SetScale{0.4}
  \SetWidth{1}
\Line(-25,30)(-40,30)
\Line(-25,-30)(-40,-30)
\Line(-25,-30)(25,-30)
\Line(-25,30)(-25,-30)
\Line(-25,-30)(25,30)
\Line(25,-10)(-25,30)
%
  \SetWidth{4}
\Line(25,-30)(40,-30)
\Line(25,30)(40,30)
\Line(25,-30)(25,30)
%
%
%
\Text(0,-26)[c]{\footnotesize{(${\mathcal T}^{A}_{44},{\mathcal T}^{B}_{38}$)}}
%
\end{picture}}
}
\hspace{1.75cm}
\vcenter{
\hbox{
  \begin{picture}(0,0)(0,0)
\SetScale{0.4}
  \SetWidth{1}
\Line(-25,30)(-40,30)
\Line(-25,-30)(-40,-30)
\Line(-25,-30)(25,-30)
\Line(-25,30)(-25,-30)
\Line(-25,-30)(25,30)
\Line(25,-10)(-25,30)
%
  \SetWidth{4}
\Line(25,-30)(40,-30)
\Line(25,30)(40,30)
\Line(25,-30)(25,30)
\Text(3,20)[c]{\tiny{$(p_1\!\!+\!\!k_1\!\!+\!\!k_2)^2$}}
\Text(0,-26)[c]{\footnotesize{(${\mathcal T}^{A}_{45},{\mathcal T}^{B}_{39}$)}}
%
\end{picture}}
}
\hspace{1.75cm}
\vcenter{
\hbox{
  \begin{picture}(0,0)(0,0)
\SetScale{0.4}
  \SetWidth{1}
\Line(-25,30)(-40,30)
\Line(-25,-30)(-40,-30)
\Line(-25,-30)(25,-30)
\Line(-25,30)(-25,-30)
\Line(-25,-30)(25,30)
\Line(25,-10)(-25,30)
%
  \SetWidth{4}
\Line(25,-30)(40,-30)
\Line(25,30)(40,30)
\Line(25,-30)(25,30)
%
%
%
\Text(3,20)[c]{\tiny{$(p_1\!\!+\!\!p_2\!\!+\!\!k_1)^2$}}
\Text(0,-26)[c]{\footnotesize{(${\mathcal T}^{A}_{46},{\mathcal T}^{B}_{40}$)}}
%
\end{picture}}
}
\hspace{1.75cm}
\vcenter{
\hbox{
  \begin{picture}(0,0)(0,0)
\SetScale{0.4}
  \SetWidth{1}
\Line(-25,30)(-40,30)
\Line(-25,-30)(-40,-30)
\Line(-25,-30)(25,-30)
\Line(-25,30)(-25,-30)
\Line(-25,-30)(25,30)
\Line(25,-10)(-25,30)
%
  \SetWidth{4}
\Line(25,-30)(40,-30)
\Line(25,30)(40,30)
\Line(25,-30)(25,30)
\Text(3,20)[c]{\tiny{$(p_1\!\!+\!\!p_2\!\!+\!\!k_2)^2$}}
\Text(0,-26)[c]{\footnotesize{(${\mathcal T}^{A}_{47},{\mathcal T}^{B}_{41}$)}}
%
\end{picture}}
}
\hspace{1.60cm}
\vcenter{
\hbox{
  \begin{picture}(0,0)(0,0)
\SetScale{0.4}
  \SetWidth{1.0}
\DashLine(-50,0)(-35,0){3}
\Line(-35,0)(10,-30)
\Line(10,30)(-11,-16)
\Line(10,-30)(25,-30)
\Line(10,-30)(-11,16)
  \SetWidth{4}
\Line(10,30)(25,30)
\Line(10,30)(-35,0)
\Text(-18,8)[c]{\footnotesize{$t$}}
\Text(0,-26)[c]{\footnotesize{(${\mathcal T}^{B}_{35}$)}}
\end{picture}}
}
\hspace{1.50cm}
\vcenter{
\hbox{
  \begin{picture}(0,0)(0,0)
\SetScale{0.4}
  \SetWidth{1}
\Line(-25,30)(-40,30)
\Line(-25,-30)(-40,-30)
\Line(-25,-30)(25,-30)
\Line(-25,30)(-25,-30)
\Line(-25,30)(25,10)
\Line(25,30)(-25,30)
%
  \SetWidth{4}
\Line(25,-30)(40,-30)
\Line(25,30)(40,30)
\Line(25,-30)(25,30)
%
%
%
\Text(0,-26)[c]{\footnotesize{(${\mathcal T}^{B}_{36}$)}}
%
\end{picture}}
}
\hspace{1.75cm}
\vcenter{
\hbox{
  \begin{picture}(0,0)(0,0)
\SetScale{0.4}
  \SetWidth{1}
\Line(-25,30)(-40,30)
\Line(-25,-30)(-40,-30)
\Line(-25,-30)(25,-30)
\Line(-25,30)(-25,-30)
\Line(-25,30)(25,10)
\Line(25,30)(-25,30)
%
  \SetWidth{4}
\Line(25,-30)(40,-30)
\Line(25,30)(40,30)
\Line(25,-30)(25,30)
\Text(3,20)[c]{\tiny{$(p_1\!\!+\!\!p_2\!\!+\!\!k_2)^2$}}
\Text(0,-26)[c]{\footnotesize{(${\mathcal T}^{B}_{37}$)}}
%
\end{picture}}
}
\]
\vspace*{0.9cm}
\[
\vcenter{
\hbox{
  \begin{picture}(0,0)(0,0)
\SetScale{0.4}
  \SetWidth{1}
\Line(-25,30)(-40,30)
\Line(-25,-30)(-40,-30)
\Line(-25,-30)(25,-30)
\Line(-25,30)(-25,-30)
\Line(-5,-30)(25,30)
\Line(25,-10)(-25,30)
%
  \SetWidth{4}
\Line(25,-30)(40,-30)
\Line(25,30)(40,30)
\Line(25,-30)(25,30)
%
%
%
\Text(0,-26)[c]{\footnotesize{(${\mathcal T}^{A}_{49}$)}}
%
\end{picture}}
}
\hspace{1.55cm}
%
\vcenter{
\hbox{
  \begin{picture}(0,0)(0,0)
\SetScale{0.4}
  \SetWidth{1}
\Line(-25,30)(-40,30)
\Line(-25,-30)(-40,-30)
\Line(-25,-30)(25,-30)
\Line(-25,30)(-25,-30)
\Line(-5,-30)(25,30)
\Line(25,-10)(-25,30)
%
  \SetWidth{4}
\Line(25,-30)(40,-30)
\Line(25,30)(40,30)
\Line(25,-30)(25,30)
%
%
%
\Text(3,20)[c]{\tiny{$(p_1\!\!+\!\!k_1\!\!+\!\!k_2)^2$}}
\Text(0,-26)[c]{\footnotesize{(${\mathcal T}^{A}_{50}$)}}
%
\end{picture}}
}
\hspace{1.55cm}
%
\vcenter{
\hbox{
  \begin{picture}(0,0)(0,0)
\SetScale{0.4}
  \SetWidth{1}
\Line(-25,30)(-40,30)
\Line(-25,-30)(-40,-30)
\Line(-25,-30)(25,-30)
\Line(-25,30)(-25,-30)
\Line(-5,-30)(25,30)
\Line(25,-10)(-25,30)
%
  \SetWidth{4}
\Line(25,-30)(40,-30)
\Line(25,30)(40,30)
\Line(25,-30)(25,30)
%
%
%
\Text(3,20)[c]{\tiny{$(p_1\!\!+\!\!p_2\!\!+\!\!k_1)^2$}}
\Text(0,-26)[c]{\footnotesize{(${\mathcal T}^{A}_{51}$)}}
%
\end{picture}}
}
\hspace{2.3cm}
%
\vcenter{
\hbox{
  \begin{picture}(0,0)(0,0)
\SetScale{0.4}
  \SetWidth{1}
\Line(-25,30)(-40,30)
\Line(-25,-30)(-40,-30)
\Line(-25,-30)(25,-30)
\Line(-25,30)(-25,-30)
\Line(-5,-30)(25,30)
\Line(25,-10)(-25,30)
%
  \SetWidth{4}
\Line(25,-30)(40,-30)
\Line(25,30)(40,30)
\Line(25,-30)(25,30)
%
%
%
\Text(3,20)[c]{\tiny{$(p_1\!\!+\!\!p_2\!\!+\!\!k_1)^2(p_1\!\!+\!\!p_2\!\!+\!\!k_2)^2$}}
\Text(0,-26)[c]{\footnotesize{(${\mathcal T}^{A}_{52}$)}}
%
\end{picture}}
}
\hspace{2.3cm}
\vcenter{
\hbox{
  \begin{picture}(0,0)(0,0)
\SetScale{0.4}
  \SetWidth{1}
\Line(-25,30)(-40,30)
\Line(-25,-30)(-40,-30)
\Line(-25,-30)(25,-30)
\Line(-25,30)(-25,-30)
\Line(-25,-10)(25,30)
\Line(25,-10)(-25,30)
%
  \SetWidth{4}
\Line(25,-30)(40,-30)
\Line(25,30)(40,30)
\Line(25,-30)(25,30)
%
%
%
\Text(0,-26)[c]{\footnotesize{(${\mathcal T}^{B}_{42}$)}}
%
\end{picture}}
}
\hspace{1.55cm}
%
\vcenter{
\hbox{
  \begin{picture}(0,0)(0,0)
\SetScale{0.4}
  \SetWidth{1}
\Line(-25,30)(-40,30)
\Line(-25,-30)(-40,-30)
\Line(-25,-30)(25,-30)
\Line(-25,30)(-25,-30)
\Line(-25,-10)(25,30)
\Line(25,-10)(-25,30)
%
  \SetWidth{4}
\Line(25,-30)(40,-30)
\Line(25,30)(40,30)
\Line(25,-30)(25,30)
\Text(3,20)[c]{\tiny{$(p_1\!\!+\!\!p_2\!\!+\!\!k_1)^2$}}
\Text(0,-26)[c]{\footnotesize{(${\mathcal T}^{B}_{43}$)}}
%
\end{picture}}
}
\hspace{1.55cm}
\vcenter{
\hbox{
  \begin{picture}(0,0)(0,0)
\SetScale{0.4}
  \SetWidth{1}
\Line(-25,30)(-40,30)
\Line(-25,-30)(-40,-30)
\Line(-25,-30)(25,-30)
\Line(-25,30)(-25,-30)
\Line(-25,-10)(25,30)
\Line(25,-10)(-25,30)
%
  \SetWidth{4}
\Line(25,-30)(40,-30)
\Line(25,30)(40,30)
\Line(25,-30)(25,30)
\Text(3,20)[c]{\tiny{$(p_1\!\!+\!\!p_2\!\!+\!\!k_2)^2$}}
\Text(0,-26)[c]{\footnotesize{(${\mathcal T}^{B}_{44}$)}}
%
\end{picture}}
}
\]
\vspace*{6mm}
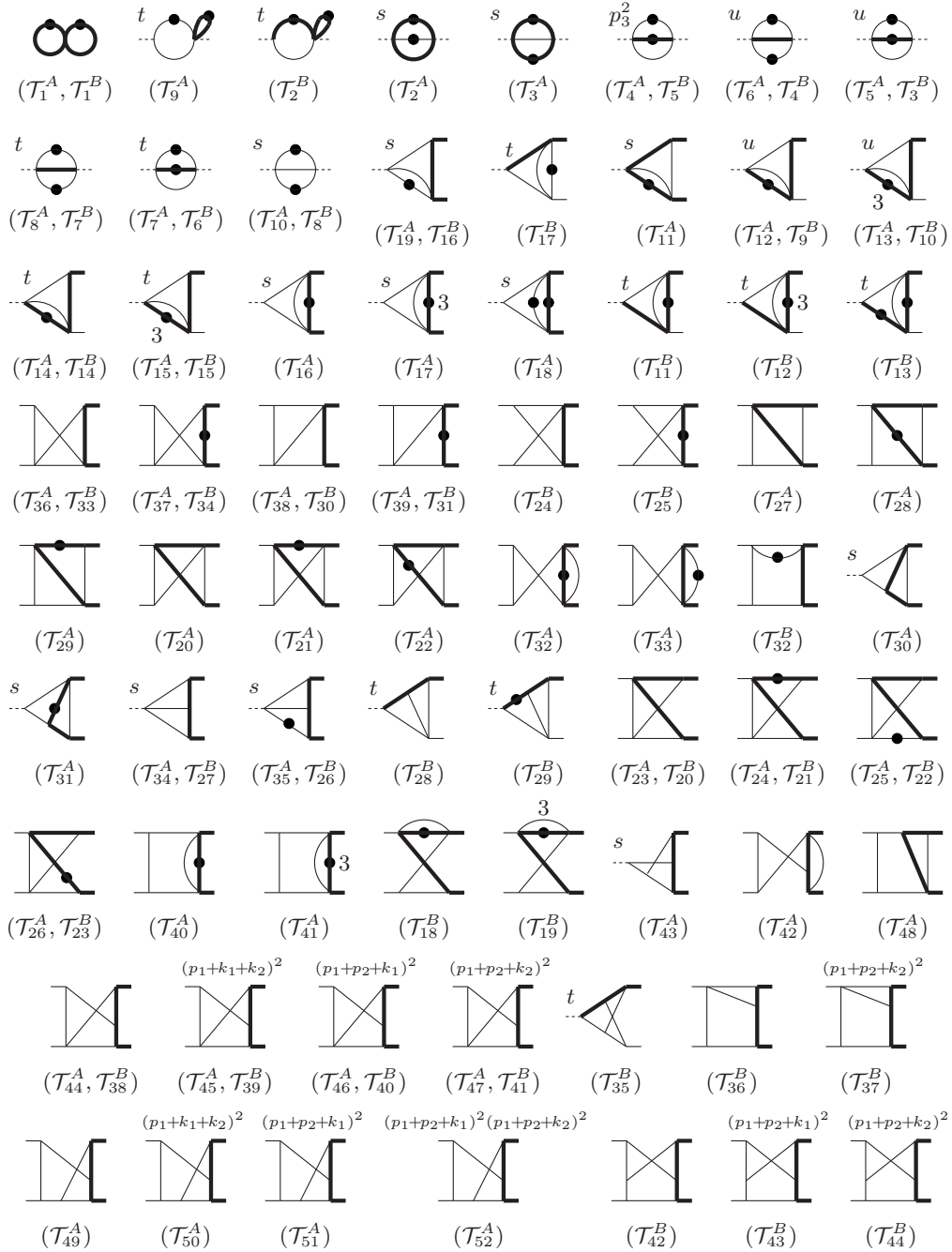
\caption{Master Integrals in pre-canonical form. Internal thin lines represent massless propagators, while thick lines represent  heavy-quark (massive) propagators. External  thin lines represent massless particles on their mass-shell, $p^2=0$. External thick lines represent massive particles on their mass-shell, $p^2=m^2$. 
\label{fig2}}
\ec
\efig
%

In this Section, we present the canonical basis used in this work. In particular, we provide the relations  that allow one to go from  the MIs in pre-canonical form (see Fig.~\ref{fig2}) to MIs in canonical form,
where the latter satisfy a differential equation of the form \eqref{CanSys}. These relations are written assuming that the Mandelstam invariants $s$ and $t$ take values in the physical region of phase space.
The normalizing prefactors contain two square roots, one of which enters only through the two integrals $f^A_{44}$ and $f^B_{38}$, see below. As will be shown in the next section, it is possible to rationalize these roots by a suitable reparametrization.
We will extend the definition of the canonical MIs also to other phase-space regions
by using the expressions listed in (\ref{eq:4.1} -- \ref{eq:4.96}) \emph{after rationalization}.
In other words, we effectively define our canonical MIs using rational prefactors in
the  parameters \{$w$, $z$\} rather than analytically continue root-valued prefactors in the original Mandelstam invariants.

The pre-canonical MIs basis is shown graphically   in Fig.~\ref{fig2}. As discussed in the caption,
thin lines represent massless propagators and external legs, while massive lines represent massive propagators and external legs. For two and three point functions, we indicate explicitly  the Mandelstam variable on which a given MI depends by adding  ``$s$'', ``$t$'', ``$u$''  or ``$p_3^2$'' to the drawing (see for example ${\mathcal T}^{A}_{9}$,${\mathcal T}^{B}_{2}$, etc.). For sub-topologies involving several MIs, dotted propagators indicate a squared propagator in the integrand of the MI (see for example ${\mathcal T}^{B}_{17}$). A dotted propagator with a $3$ next to the dot indicates a cubed propagator in the integrand (see for example ${\mathcal T}^{B}_{12}$). The four-point subtopologies in the last two lines of Fig.~\ref{fig2} involve several MIs which differ from one another because of the numerator in the integrand. These numerators are shown on top of the drawing of each single MI (see for example ${\mathcal T}^{B}_{37}$, etc). In order to avoid any possible misinterpretation of Fig.~\ref{fig2}, the  integral definition of each MIs in the pre-canonical  basis  can be found  in Appendix~\ref{pre-can}.
In addition, we provide all definitions in the ancillary files of the {\tt arXiv} submission of this work.

\subsection{Topology $A$}

Topology $A$ involves 52 MIs. Their canonical form is obtained with the following change of basis:

\begin{eqnarray}
f^{A}_{1}&=& \epsilon^2\,\T^{A}_{1}\, , \label{eq:4.1} \\
f^{A}_{2}&=& \epsilon^2 \sqrt{s(s-4 m^2)} \, \T^{A}_{2}\,+\,\frac{1}{2} \epsilon^2 \sqrt{s(s-4m^2)} \T^{A}_{3}\, , \\
f^{A}_{3}&=& \epsilon^2\,s\,\T^{A}_{3}\, ,  \\
f^{A}_{4}&=& \epsilon^2\,m^2\,\T^{A}_{4} \, ,\\
f^{A}_{5}&=& \epsilon ^2 \left(-2 m^2+s+t\right)\,\T^{A}_{5}\, , \\
f^{A}_{6}&=& \epsilon ^2 \left(s+t-m^2\right)\,\T^{A}_{6}\,+\,2\,\epsilon ^2\, m^2\,\T^{A}_{5}\, , \\
f^{A}_{7}&=& \epsilon^2\,t\,\T^{A}_{7}\, , \\
f^{A}_{8}&=& - \epsilon ^2 \left(m^2-t\right)\,\T^{A}_{8}\,-2\, \epsilon ^2\, m^2\,\T^{A}_{7} \, ,\\
f^{A}_{9}&=& \epsilon^2\,s\,\T^{A}_{9}\, , \\
f^{A}_{10}&=& \epsilon^2\,s\,\T^{A}_{10}\, , \\
f^{A}_{11}&=& \epsilon ^3 \sqrt{s(s-4 m^2)}\,\T^{A}_{11}\, , \\
f^{A}_{12}&=& \epsilon ^3 \left(s+t-m^2\right)\,\T^{A}_{12}\, , \\
f^{A}_{13}&=& \epsilon ^2 \,m^2\, \left(s+t-m^2\right)\,\T^{A}_{13}\, , \\
f^{A}_{14}&=& - \epsilon ^3\, \left(m^2-t\right)\,\T^{A}_{14} \, ,\\
f^{A}_{15}&=& - \epsilon ^2 \,m^2\, \left(m^2-t\right)\,\T^{A}_{15} \, ,\\
f^{A}_{16}&=& \epsilon ^3 \sqrt{s(s-4 m^2)}\,\T^{A}_{16}\, , \\
f^{A}_{17}&=& \epsilon ^2 \,m^2\,\sqrt{s(s-4 m^2)}\,\T^{A}_{17}\, , \\
f^{A}_{18}&=& \epsilon ^2\, m^2 \,s\,\T^{A}_{18} + \epsilon^2 4 m^2 s \T^{A}_{17}  -3 \epsilon^3 s \T^{A}_{16}\, ,\\
f^{A}_{19}&=& \epsilon ^3\, \sqrt{s(s-4 m^2)}\,\T^{A}_{19}\, , \\
f^{A}_{20}&=& \epsilon ^4 \,\left(m^2-t\right)\,\T^{A}_{20}\, , \\
f^{A}_{21}&=& \epsilon ^3\,\sqrt{s(s-4 m^2)}\, \left(s+t-m^2\right)\,\T^{A}_{21}\, , \\
f^{A}_{22}&=& \epsilon ^3 \,m^2\, \left(s+t-m^2\right)\,\T^{A}_{22}\,+\,\frac{1}{2} \epsilon ^3 \left(2 m^2-s\right) \left(s+t-m^2\right)\,\T^{A}_{21}\, , \\
f^{A}_{23}&=& \epsilon ^4 \sqrt{s(s-4 m^2)}\,\T^{A}_{23}\, ,  \\
f^{A}_{24}&=& - \epsilon ^3 \, m^2 \,\left(m^2-t\right)\,\T^{A}_{24}\, ,  \\
f^{A}_{25}&=& - \epsilon ^3 \, \left(m^2-t\right) \left(s+t-m^2\right)\,\T^{A}_{25}\, ,  \\
f^{A}_{26}&=& \epsilon ^3 \,m^2\, \left(s+t-m^2\right)\,\T^{A}_{26} \, , \\
f^{A}_{27}&=& \epsilon ^4 \,\left(s+t-m^2\right)\,\T^{A}_{27}\, ,  \\
f^{A}_{28}&=& - \epsilon ^3 m^2 \left(m^2-t\right)\,\T^{A}_{28}\, + \frac{1}{2} \epsilon ^3 \left(s-2 m^2\right) \left(m^2-t\right) \,\T^{A}_{29}\, ,  \\
f^{A}_{29}&=& - \epsilon ^3 \, \sqrt{s(s-4 m^2)} \left(m^2-t\right)\,\T^{A}_{29}\, ,  \\
f^{A}_{30}&=& \epsilon ^4 \sqrt{s(s-4 m^2)}\,\T^{A}_{30} \, , \\
f^{A}_{31}&=& \epsilon ^2\, m^2\, s\,\T^{A}_{31}\,-2 \epsilon ^4 \,s\,\T^{A}_{30}\,+\,\epsilon ^3 s\,\T^{A}_{11}\,+\,\epsilon ^3 s\,\T^{A}_{16}\,-2 \epsilon ^2 m^2 s\,\T^{A}_{17} \, , \\
f^{A}_{32}&=& \epsilon ^3 \,s\, \left(s+t-2 m^2\right)\,\T^{A}_{32}\, ,  \\
f^{A}_{33}&=& \epsilon ^3 \,s\, \left(s+t-m^2\right)\,\T^{A}_{33}\,+\,\epsilon ^3\, m^2\, s\,\T^{A}_{32}\, ,  \\
f^{A}_{34}&=& \epsilon ^4 \,\sqrt{s(s-4 m^2)}\,\T^{A}_{34} \, , \\
f^{A}_{35}&=& \epsilon ^2 (2 \epsilon +1) \,m^2 \,s\,\T^{A}_{35}\,+\,2 \epsilon ^4 \,s\,\T^{A}_{34}\,-\frac{\epsilon ^3 s}{2}\,\T^{A}_{19}\, ,  \\
f^{A}_{36}&=& - \epsilon ^4\, \left(m^2-t\right)\,\T^{A}_{36}\, ,  \\
f^{A}_{37}&=& \epsilon ^3\, m^2\, s\,\T^{A}_{37}\, ,  \\
f^{A}_{38}&=& \epsilon ^4 \,\left(s+t-m^2\right)\,\T^{A}_{38}\, ,  \\
f^{A}_{39}&=& \epsilon ^3 \,m^2\, s\,\T^{A}_{39}\, ,  \\
f^{A}_{40}&=& \epsilon ^3 \,s\, t\,\T^{A}_{40}\, ,  \\
f^{A}_{41}&=& - \epsilon ^2 \,m^2\, s\, \left(m^2-t\right)\,\T^{A}_{41}\,+\, \epsilon ^3 \,m^2\, s\, \T^{A}_{40}\, ,  \\
f^{A}_{42}&=& \epsilon ^4 \,s\, \left(s+t-m^2\right)\,\T^{A}_{42}\, ,  \\
f^{A}_{43}&=& \epsilon ^4 \,s\, \sqrt{s(s-4 m^2)}\,\T^{A}_{43}\, ,  \\
f^{A}_{44}&=& \epsilon ^4 \, i\sqrt{m^2\, s(m^2-t)(s+t-m^2)}\,\T^{A}_{44} \, , \\
f^{A}_{45}&=& \epsilon ^4 \sqrt{s(s-4 m^2)}\,\T^{A}_{45}\,+\,\epsilon ^4 \sqrt{s(s-4 m^2)} \left(s+t-m^2\right)\,\T^{A}_{44}\nn\\
&&+\,\epsilon ^4 \sqrt{s(s-4 m^2)}\,\T^{A}_{46}\,+\,\epsilon ^4 
\sqrt{s(s-4 m^2)}\,\T^{A}_{47}\,+\frac{\epsilon ^2 m^2 s \sqrt{s(s-4 m^2)}}{\left(m^2-t\right) \left(s+t-m^2\right)}\,\T^{A}_{4}\nn\\
&&+\frac{\epsilon ^2 s \left(12 m^4 -m^2 (7 s+4 t)+s (s+t)\right)}{2 \sqrt{s(s-4 m^2)} \left(s+t-m^2\right)}\,\T^{A}_{5}\nn\\
&&-\frac{1}{4} \epsilon ^2 \sqrt{s(s-4 m^2)}\,\T^{A}_{6}\,
-\frac{\epsilon ^2 \sqrt{s(s-4 m^2)} \left(m^2+t\right)}{2 \left(m^2-t\right)}\,\T^{A}_{7}\nn\\
&&-\frac{1}{4} \epsilon ^2 \sqrt{s(s-4 m^2)}\,\T^{A}_{8}\,+\epsilon ^3 \sqrt{s(s-4 m^2)}\,\T^{A}_{12}\,-\epsilon ^2 m^2 \sqrt{s(s-4 m^2)}\,\T^{A}_{13}\nn\\
&&+\epsilon ^3 \sqrt{s(s-4 m^2)}\,\T^{A}_{14}\,-\epsilon ^2 m^2 \sqrt{s(s-4 m^2)}\,\T^{A}_{15} \, ,\\
f^{A}_{46}&=& \epsilon ^4 \left(s+t-m^2\right)\,\T^{A}_{46}\,+\,\epsilon ^4 s \left(s+t-m^2\right)\,\T^{A}_{44}\,+\,\epsilon ^4\,s\,\T^{A}_{47}\nn\\
&&+\epsilon ^2 \left(\frac{m^2 s}{m^2-t}+m^2\right)\,\T^{A}_{4}\,-\,\frac{1}{2} \epsilon ^2 \left(\frac{2 m^2 s}{m^2-t}-s+t\right)\,\T^{A}_{7}\,-\frac{\epsilon ^2 s}{4}\,\T^{A}_{8} \nn\\
&&+\,\epsilon ^3\,s\,\T^{A}_{14}\,-\epsilon ^2 \,m^2 \,s\,\T^{A}_{15}\,-\epsilon ^4 \left(s+t-m^2\right)\,\T^{A}_{23}\,+\,\epsilon ^4 s\,\T^{A}_{36}\nn\\
&&-\frac{1}{2} \epsilon ^4 \left(s-t+m^2\right) \,\T^{A}_{38} \, , \\
f^{A}_{47}&=& \epsilon ^4 \left(t-m^2\right)\,\T^{A}_{47}\,+\frac{1}{2} \epsilon ^4 \left(s-t+m^2\right) \,\T^{A}_{38} \, , \\
f^{A}_{48}&=& - \epsilon ^4 \,s \left(m^2-t\right)\,\T^{A}_{48} \, , \\
f^{A}_{49}&=& \epsilon ^4 \,s^2 \left(s+t-m^2\right) \, \T^{A}_{49}\,+\,\epsilon ^4\, s^2\,\T^{A}_{51}\nn\\
&&+\epsilon ^2 \left( - \frac{3 m^2 s}{2 \left(m^2-t\right)}+\frac{3 m^2 s}{2 \left(s+t-m^2\right)}-\frac{9 m^2}{4}\right)\,\T^{A}_{4} \nn\\
&&+\epsilon ^2 \left(-\frac{3 m^2 s}{2 \left(s+t-m^2\right)}+m^2+\frac{s}{2}-\frac{t}{2}\right)\,\T^{A}_{5}\,-\frac{\epsilon ^2 s}{4}\,\T^{A}_{6}\nn\\
&&+\epsilon ^2 \left(\frac{m^2 s}{2 \left(m^2-t\right)}+\frac{s t}{m^2-t}\right)\,\T^{A}_{7}\,+\,\frac{\epsilon ^2 s}{4}\,\T^{A}_{8}\nn\\
&&+\epsilon ^2 \left(\frac{3 s^2}{4 \left(m^2-t\right)}-\frac{3 s^2}{4 \left(s+t-m^2\right)}+\frac{9 s}{8}\right)\,\T^{A}_{10}+\frac{ \epsilon^4 s \left(m^2-t\right)}{1+4 \epsilon}\,\T^A_{21}
\nn\\
&&
+\frac{ \epsilon^4 s \left(m^2-t\right)}{1+4 \epsilon}\,\T^A_{25}
+\frac{ \epsilon^4 s \left(m^2-t\right)}{1+4 \epsilon}\,\T^A_{29}
\nn\\
&&-\frac{1}{4} \epsilon ^3 s \left(s-t+m^2\right)\,\T^{A}_{32}-\frac{1}{4} \epsilon ^3 s \left(s-t+m^2\right)\,\T^{A}_{33}\nn\\
&&+\frac{1}{2} \epsilon ^4 \left(m^2-t\right) \left(\frac{6 s}{s+t-m^2}-3\right)\,\T^{A}_{36}+\epsilon ^3 s \left(m^2-\frac{2 m^2 s}{s+t-m^2}\right)\,\T^{A}_{37}\nn\\
&&-\frac{3 \epsilon ^4 \left(s+t-m^2\right) \left(s-t+m^2\right)}{m^2-t}\,\T^{A}_{38}\,
+\frac{2 \epsilon ^3 m^2 s \left(s-t+m^2\right)}{m^2-t}\,\T^{A}_{39}\,+\epsilon ^3 s^2\,\T^{A}_{40}\nn\\
&&-\epsilon ^2 m^2 s^2\,\T^{A}_{41}\, +\epsilon ^4 s \left(s+t-m^2\right)\,\T^{A}_{44}\,-\epsilon ^4 s^2\,\T^{A}_{48} \, , \\
f^{A}_{50}&=& \epsilon ^4 s \sqrt{s(s-4 m^2)}\,\T^{A}_{50}\,-\epsilon ^4 s \sqrt{s(s-4 m^2)}\,\T^{A}_{42} + \epsilon ^4 \sqrt{s(s-4 m^2)} \left(m^2-t\right)\,\T^{A}_{44}  \, ,
 \\
f^{A}_{51}&=& - \epsilon ^4 s \left(m^2-t\right)\,\T^{A}_{51}\,+\epsilon ^2 \left(\frac{3 m^2}{4}-\frac{3 m^2 s}{2 \left(s+t-m^2\right)}\right)\,\T^{A}_{4}\nn\\
&&+\frac{1}{2} \epsilon ^2 \left(m^2 \left(\frac{3 s}{s+t-m^2}-2\right)-s+t\right)\,\T^{A}_{5}\,+\frac{\epsilon ^2 s}{4}\,\T^{A}_{6}\,+\frac{3 \epsilon ^2 s \left(s-t+m^2\right)}{8 \left(s+t-m^2\right)}\,\T^{A}_{10}\nn\\
&&
-\frac{\epsilon ^4 s\left(m^2-t\right)}{1+4 \epsilon} \,\T^{A}_{21} 
-\frac{\epsilon ^4 s\left(m^2-t\right)}{1+4 \epsilon} \,\T^{A}_{25} 
\nn\\ 
&&
-\frac{\epsilon ^4 s\left(m^2-t\right)}{1+4 \epsilon} \,\T^{A}_{29} 
\,+\frac{1}{4} \epsilon ^3 s \left(s-t+m^2\right)\,\T^{A}_{32}\nn\\
&&+\frac{1}{4} \epsilon ^3 s \left(s-t+m^2\right)\,\T^{A}_{33}\,+\frac{1}{2} \epsilon ^4 \left(m^2-t\right) \left(3-\frac{6 s}{s+t-m^2}\right)\,\T^{A}_{36}\nn\\
&&+\epsilon ^3 m^2 s \left(\frac{2 s}{s+t-m^2}-1\right)\,\T^{A}_{37}  \, ,\\
f^{A}_{52}&=& \epsilon ^4 \,s\,\T^{A}_{52}\,+\,\frac{\epsilon ^4 s^2}{2}\,\T^{A}_{50}\,
+\frac{4 \epsilon^5}{1-4 \epsilon^2}\,\T^{A}_{1} 
\nn\\&&
+\epsilon ^2 \left(\epsilon  \left(s-4 m^2\right)+\frac{s-4 m^2}{2 (2 \epsilon -1)}-2 m^2+\frac{s}{2}\right)\,\T^{A}_{2}\nn\\
&&+\epsilon ^2 \left(\frac{-4 m^2-s}{4 (2 \epsilon -1)}+\epsilon  \left(-2 m^2-s\right)-m^2\right)\,\T^{A}_{3}\nn\\
&&+\epsilon ^2 \left(\frac{3 m^2}{2 (2 \epsilon -1)}+\frac{m^2 s}{4 \left(m^2-t\right)}+\frac{m^2 s}{2
\left(s+t-m^2\right)}\right)\,\T^{A}_{4}\nn\\
&&+\epsilon ^2 \left(-\frac{m^2 s}{2 \left(s+t-m^2\right)}+\frac{m^2}{4}+\frac{s}{8}-\frac{t}{8}\right)\,\T^{A}_{5}\,-\frac{\epsilon ^2 s}{8}\,\T^{A}_{6}\nn\\
&&+\epsilon ^2 \left(\frac{-2 m^2-t}{2 (2 \epsilon -1)}-\frac{\left(m^2+t\right) \left(s-t+m^2\right)}{8\left(m^2-t\right)}\right)\,\T^{A}_{7} \nn \\
&&
+\epsilon ^2 \left(\frac{t-m^2}{2 (2 \epsilon -1)}-\frac{m^2}{16}-\frac{s}{16}+\frac{t}{16}\right)\,\T^{A}_{8}
+\epsilon ^3 s\,\T^{A}_{9}
-\,\frac{\epsilon^4 s}{2 \epsilon -1}\,\T^{A}_{11}
+\frac{\epsilon ^3 s}{2}\,\T^{A}_{12}\, \nn \\
&&-\frac{1}{4} \epsilon ^2 m^2 \left(s-t+m^2\right)\,\T^{A}_{13}+\epsilon ^3 \left(\frac{3 \left(m^2-t\right)}{2 (2 \epsilon -1)}
+\frac{m^2}{4}+\frac{s}{4}-\frac{t}{4}\right)\,\T^{A}_{14}
\nn \\ &&
+\epsilon ^2 \left(-\frac{2 m^2 \left(m^2-t\right)}{2 \epsilon -1}-\frac{1}{4} m^2 \left(s-t+m^2\right)\right)\,\T^{A}_{15}
+\,\frac{\epsilon ^4 s}{2 \epsilon-1}\, \T^{A}_{16}\,
-\, \frac{ 2 \epsilon ^3 m^2 s}{2 \epsilon-1}\, \T^{A}_{17} \nn \\ 
&&
-\, \frac{ 4 \epsilon ^4 s}{2 \epsilon-1}\, \T^{A}_{19}-\frac{1}{8} \epsilon ^3 s \left(s+t-m^2\right)\,\T^{A}_{21}\,+\epsilon ^4 \left(\frac{t}{4}-\frac{m^2}{4}\right)\,\T^{A}_{23}
\nn\\
&&+\,\epsilon ^4 \left(\frac{3 m^2-2 s-3 t}{2 \epsilon -1}+s\right)\,\T^{A}_{27}-\frac{\epsilon ^3 m^2 \left(m^2-t\right)}{2 \epsilon -1}\,\T^{A}_{28}
\nn\\
&&+\epsilon ^3 \left(\frac{1}{8} s \left(m^2-t\right)-\frac{\left(2 m^2-s\right) \left(m^2-t\right)}{2 (2 \epsilon -1)}\right)\,\T^{A}_{29}\,+\epsilon ^3 m^2 s\,\T^{A}_{31}\nn\\
&&+\epsilon ^3 \left(\frac{s \left(s+t-m^2\right)}{2 (2 \epsilon -1)}+\frac{m^2 s}{2}\right)\,\T^{A}_{32}\,-\frac{\epsilon ^4 s}{4}\,\T^{A}_{34}\,+\epsilon ^4 \left(-\frac{s}{2 \epsilon -1}-\frac{5}{4} \left(s-t+m^2\right)\right)\,\T^{A}_{36}\nn\\
&&- \epsilon ^4 \left(\frac{1}{2} \left(s-t+m^2\right)+\frac{s}{2 \epsilon -1}\right)\,\T^{A}_{38}\,
-\epsilon ^4 \frac{s}{2}\,\T^{A}_{42} 
+\frac{\epsilon ^4 s^2}{4}\,\T^{A}_{43}\,+\frac{1}{4} \epsilon ^4 s \left(s-t+m^2\right)\,\T^{A}_{44}\nn\\
&&
+\frac{\epsilon ^4 s}{2}\,\T^{A}_{45}\,+\frac{1}{4} \epsilon ^4 \left(s-t+m^2\right)\,\T^{A}_{46}\,+\frac{1}{4} \epsilon ^4 \left(s-t+m^2\right)\,\T^{A}_{47} \, .
\end{eqnarray}

\subsection{Topology $B$}

Topology $B$ involves 44 MIs. The relations linking the canonical and pre-canonical forms of the MI basis are the following:

\begin{eqnarray}
f^{B}_{1}&=& \epsilon^2\, \T^{B}_{1}\, , \\
f^{B}_{2}&=& \epsilon^2 t\,\T^{B}_{2}\, , \\
f^{B}_{3}&=& \epsilon ^2 \left(s+t-2 m^2\right)\,\T^{B}_{3} \, ,\\
f^{B}_{4}&=& \epsilon ^2 \left(s+t-m^2\right)\,\T^{B}_{4}\,+\,2\,\epsilon^2 \,m^2\,\T^{B}_{3}\, , \\
f^{B}_{5}&=& \epsilon^2 m^2 \,\T^{B}_{5} \, ,\\
f^{B}_{6}&=& \epsilon^2\,t\,\T^{B}_{6} \, ,\\
f^{B}_{7}&=& - \epsilon^2 \left(m^2-t\right)\,\T^{B}_{7}\,-\,2\,\epsilon^2\,m^2\,\T^{B}_{6}\, , \\
f^{B}_{8}&=& \epsilon^2 \,s\,\T^{B}_{8} \, ,\\
f^{B}_{9}&=& - \epsilon ^3 \left(s+t-m^2\right)\,\T^{B}_{9}\, , \\
f^{B}_{10}&=& - \epsilon ^2\,m^2\, \left(s+t-m^2\right)\,\T^{B}_{10} \, ,\\
f^{B}_{11}&=& \epsilon ^3 \left(m^2-t\right)\,\T^{B}_{11} \, ,\\
f^{B}_{12}&=& \epsilon ^2\,m^2\, \left(m^2-t\right)\,\T^{B}_{12} \, ,\\
f^{B}_{13}&=& \epsilon ^2\,m^2\, \left(2 m^2-t\right)\,\T^{B}_{13}\,-3\,\epsilon^3\,m^2\,\T^{B}_{11}\,+\,2\,\epsilon^2\,m^4\,\T^{B}_{12} \, ,\\
f^{B}_{14}&=& \epsilon ^3 \left(m^2-t\right)\,\T^{B}_{14}\, , \\
f^{B}_{15}&=& \epsilon^2\,m^2\, \left(m^2-t\right)\,\T^{B}_{15}\, , \\
f^{B}_{16}&=& \epsilon ^3\,\sqrt{s(s-4 m^2)}\,\T^{B}_{16} \, ,\\
f^{B}_{17}&=& \epsilon ^3 \left(m^2-t\right)\,\T^{B}_{17}\, , \\
f^{B}_{18}&=& \epsilon ^3 \left(m^2-t\right) \left(s+t-2 m^2\right)\,\T^{B}_{18} \\
f^{B}_{19}&=& \epsilon ^2\,m^2\, \left(m^2-t\right) \left(s+t-m^2\right)\,\T^B_{19}\,-\,\epsilon ^3 \left(m^2-t\right) \left(s+t-m^2\right)\,\T^{B}_{18}\, , \\
f^{B}_{20}&=& \epsilon ^4\,\sqrt{s(s-4 m^2)}\,\T^{B}_{20}\, , \\
f^{B}_{21}&=& \epsilon ^3 \,m^2\,\left(m^2-t\right)\,\T^{B}_{21}\, , \\
f^{B}_{22}&=& \epsilon ^3 \left(m^2-t\right) \left(s+t-m^2\right)\,\T^{B}_{22}\, , \\
f^{B}_{23}&=& \epsilon ^3\,m^2\, \left(s+t-m^2\right)\,\T^{B}_{23} \, ,\\
f^{B}_{24}&=& \epsilon^4 \, s\, \T^{B}_{24} \, ,\\
f^{B}_{25}&=& - \epsilon ^3 \left(m^2-t\right) \left(s+t-m^2\right)\,\T^{B}_{25} \, ,\\
f^{B}_{26}&=& \epsilon^2(1+2\epsilon)\,m^2\,s\,\T^{B}_{26}\,+\,2\,\epsilon^4 \, s\,\T^{B}_{27}\,-\epsilon^3\,\frac{s}{2}\,\T^{B}_{16} \, ,\\
f^{B}_{27}&=& \epsilon ^4\,\sqrt{s(s-4 m^2)}\,\T^{B}_{27}\, , \\
f^{B}_{28}&=& - \epsilon ^4 \left(m^2-t\right)\,\T^{B}_{28}\, , \\
f^{B}_{29}&=& - \epsilon ^3\,m^2 \left(m^2-t\right)\,\T^{B}_{29}\, , \\
f^{B}_{30}&=& \epsilon ^4 \left(s+t-m^2\right)\,\T^{B}_{30}\, , \\
f^{B}_{31}&=& \epsilon^3 \, m^2\,s\,\T^{B}_{31} \, ,\\
f^{B}_{32}&=& - \epsilon ^3\,s\, \left(m^2-t\right)\,\T^{B}_{32}\, , \\
f^{B}_{33}&=& - \epsilon ^4 \left(m^2-t\right)\,\T^{B}_{33} \, ,\\
f^{B}_{34}&=& \epsilon^3\,m^2\,s\,\T^{B}_{34} \, ,\\
f^{B}_{35}&=& \epsilon ^4 \left(m^2-t\right)^2\,\T^{B}_{35} \, ,\\
f^{B}_{36}&=& - \epsilon ^4\,\sqrt{s(s-4 m^2)} \left(m^2-t\right)\,\T^{B}_{36} \, ,\\
f^{B}_{37}&=& 
- \epsilon^3\,(1-2\epsilon)\left(m^2-t\right)\,\T^{B}_{37}
\,+\,2\, \epsilon ^4\,s\, \left(m^2-t\right)\,\T^{B}_{36}\,-\,2\, \epsilon ^4 \left(m^2-t\right)\,\T^{B}_{30} \, ,\\
f^{B}_{38}&=& \epsilon ^4 \, i\sqrt{m^2\, s(m^2-t)(s+t-m^2)}\,\T^{B}_{38} \, , \\
f^{B}_{39}&=& \epsilon ^4 \sqrt{s(s-4 m^2)}\,\T^{B}_{39}\,+\epsilon ^4 \sqrt{s(s-4 m^2)} \left(s+t-m^2\right)\,\T^{B}_{38}+\epsilon ^4\, \sqrt{s(s-4 m^2)}\,\T^{B}_{40}\nn\\
&&+\,\epsilon ^4\, \sqrt{s(s-4 m^2)}\,\T^{B}_{41}\,+\,\frac{\epsilon ^2\, \sqrt{s(s-4 m^2)} \left(s+t-3 m^2\right)}{2\left(s+t-m^2\right)}\,\T^{B}_{3}\nn\\
&&-\frac{\epsilon ^2\, \sqrt{s(s-4 m^2)}}{4}\,\T^{B}_{4}\,+\,\frac{\epsilon ^2\,m^2\, s\, \sqrt{s(s-4 m^2)}}{\left(m^2-t\right)\left(s+t-m^2\right)}\,\T^{B}_{5}\nn\\
&& -\frac{\epsilon ^2\, \sqrt{s(s-4 m^2)} \left(m^2+t\right)}{2 \left(m^2-t\right)}\,\T^{B}_{6}\,-\frac{\epsilon^2}{4}\, \sqrt{s(s-4 m^2)}\,\T^{B}_{7}\nn\\
&& +\epsilon ^3\,\sqrt{s(s-4 m^2)}\,\T^{B}_{9}\,-\epsilon ^2\,m^2 \sqrt{s(s-4 m^2)}\,\T^{B}_{10}
\,+\,\epsilon ^3\, \sqrt{s(s-4 m^2)}\,\T^{B}_{14}\nn\\
&&-\epsilon ^2\,m^2 \sqrt{s(s-4 m^2)}\,\T^{B}_{15}\, , \\
f^{B}_{40}&=& \epsilon^4\,s\,\T^{B}_{41}\,+\,\epsilon ^4\,s\, \left(s+t-m^2\right)\,\T^{B}_{38}\,+\,\epsilon ^4 \left(s+t-m^2\right)\,\T^{B}_{40}\,+\,\frac{ \epsilon ^2 \,m^2\, s}{m^2-t}\,\T^{B}_{5}\nn\\
&& -\frac{\epsilon ^2\,s \, \left(m^2+t\right)}{2 \left(m^2-t\right)}\,\T^{B}_{6}\,-\frac{\epsilon^2\,s}{4}\,\T^{B}_{7}\,+\,\epsilon ^3\,s\,\T^{B}_{14}\,- \epsilon ^2\,m^2\, s\,\T^{B}_{15}\,-\,\epsilon ^4 \left(s+t-m^2\right)\,\T^{B}_{20}\nn\\
&& -\epsilon^4\,s\,\T^{B}_{30}\,+\,\epsilon^4 \,s\,\T^{B}_{33}\, , \\
f^{B}_{41}&=& - \epsilon ^4 \left(m^2-t\right)\,\T^{B}_{41}\,+\,\epsilon^4\,s\,\T^{B}_{30} \, ,\\
f^{B}_{42}&=& - \epsilon ^4\,s\, \left(m^2-t\right) \left(s+t-m^2\right)\,\T^{B}_{42}\,-\,\epsilon ^4 \left(m^2-t\right) \left(s+t-m^2\right)\,\T^{B}_{43}\nn\\
&& - \epsilon ^4 \left(m^2-t\right) \left(s+t-m^2\right)\,\T^{B}_{44}\,+\,\epsilon^2\,s\,\T^{B}_{8}\,+\,\epsilon ^3\,m^2\, \left(m^2-t\right)\,\T^{B}_{18}\nn\\
&& - \, \epsilon ^2\,m^2\, \left(m^2-t\right) \left(s+t-m^2\right)\,\T^{B}_{19} \nn \\
&& - \,\epsilon ^3 \left(m^2-t\right) \left(s+t-m^2\right)\,\T^{B}_{25}\nn\\
&&+\,\epsilon^3\,m^2\,s\,\T^{B}_{31}\,-\,\epsilon ^4 \left(m^2-t\right)\,\T^{B}_{33}\,+\,\epsilon ^3\,m^2 \, s \,\T^{B}_{34}\,+\,\epsilon ^4\,s \, \left(m^2-t\right)\,\T^{B}_{35} \, , \\
f^{B}_{43}&=& \epsilon ^4 \left(m^2-t\right)^2\,\T^{B}_{43}\,+\,\epsilon^2\,s\,\T^{B}_{8}\,+\epsilon ^3 \left(2 m^4-3 m^2 t+t^2\right)\,\T^{B}_{18}\nn\\
&& - \,\epsilon ^2\,m^2\, s \, \left(m^2-t\right)\,\T^{B}_{19}\,-\frac{\epsilon ^3}{2} \left(m^2-t\right)^2\,\T^{B}_{25}\,+\,\epsilon ^4 \left(s+t-m^2\right)\,\T^{B}_{30}\nn\\
&&+\,\epsilon^3\,m^2\,s\,\T^{B}_{31}\,+\, \epsilon ^4\,s\, \left(m^2-t\right)\,\T^{B}_{38} \, , \\
f^{B}_{44}&=& \epsilon ^4 \left(m^2-t\right)^2\,\T^{B}_{44}\, \nn \\
&&+\,\epsilon^2\,\left\{
\frac{1}{4} \left[m^2 \left(\frac{3 s}{s+t-m^2}-2\right)-s+t\right]-\frac{\left(m^2-t\right) \left(5 m^2-2 (s+t)\right)}{4(1+4\epsilon)\left(s+t-m^2\right)} 
\right\} \,\T^{B}_{3}
\nn \\
&&+\epsilon ^2\,\left( - \frac{m^2-t}{8 (1+4 \epsilon)}+\frac{s}{8}\right)\,\T^{B}_{4}\nn\\
&&+\epsilon ^2 \left(\frac{3 m^2 s}{4 (1+4 \epsilon) \left(s+t-m^2\right)}-\frac{3 m^2
s}{4 \left(s+t-m^2\right)}\right)\,\T^{B}_{5}\nn\\
&&+\epsilon ^2 \left(\frac{-m^2-2 t}{4 (1+4 \epsilon)}-2 m^2\right)\,\T^{B}_{6}\,+\,\epsilon ^2 \left(- \frac{m^2-t}{8 (1+4 \epsilon )}-m^2+t\right)\,\T^{B}_{7}\nn\\
&&+\epsilon ^2 \left(\frac{3 s \left(m^2-t\right)}{8 \left(s+t-m^2\right)}-\frac{3
s^2}{8 (1+4 \epsilon) \left(s+t-m^2\right)}\right)\,\T^{B}_{8}\,+\,\epsilon ^3 \left(m^2-t\right)^2\,\T^{B}_{18}\nn\\
&&-\epsilon ^2\,m^2\, \left(m^2-t\right)^2\,\T^{B}_{19}\,-\frac{1}{2} \epsilon ^3 \left(m^2-t\right)^2\,\T^{B}_{22}\,+\,\epsilon^4\,s\,\T^{B}_{24}\nn\\
&&+\frac{\epsilon ^3}{4} \left(\frac{s \left(m^2-t\right)}{1+4 \epsilon}-
\left(m^2-t\right)^2
\right)\,\T^{B}_{25}+\,\epsilon ^4 \left(s+t-m^2\right)\left(\frac{3}{2 (1+4 \epsilon)}
+1\right)\,\T^{B}_{30}\,\nn\\
&&
+\,\frac{4 \epsilon^4 m^2 s}{1+4 \epsilon}\,\T^{B}_{31} 
-\frac{6 \epsilon^5 (m^2-t)^2}{(1+4 \epsilon)\left(s+t-m^2\right)}\,\T^{B}_{33}
\nn\\
&&+\epsilon ^3 \left(\frac{m^2 s^2}{s+t-m^2}+\frac{m^2 s
   \left(1-\frac{s}{s+t-m^2}\right)}{1+4 \epsilon}\right)\,\T^{B}_{34} \label{eq:4.96} \, .
\end{eqnarray}

\section{Rationalizing Parametrization}
 \label{sec:param}
 
Both the definition of the Canonical Basis for topologies $A$ and $B$ and the differential equations fulfilled by the MIs in the Canonical Basis contain two square roots of functions of Mandelstam invariants. These roots are:
\be 
\sqrt{s(s-4m^2)},\quad \sqrt{m^2 s (m^2-t)(s+t-m^2)},
\label{eq:roots}
\ee
where the latter root enters only through the definition of two MIs in the canonical basis: $f^A_{44}$ and $f^B_{38}$. In order to express the solution in terms of GPLs one needs  to rationalize the square roots by finding an  appropriate change of variables.
We start by defining the dimensionless variables $x$ and $y$ as follows:
\be
x = - \frac{s}{m^2} \, , \quad
y = - \frac{t}{m^2} \, .
\ee
A particularly convenient parametrization~\cite{LorenzoReparam:2018,DiVita:2018nnh} which rationalizes the roots\footnote{Another solution to the rationalization problem can be obtained  from diophantine equations as described in \cite{Becchetti:2017abb}:
$x = {16w^2(1+4z)^2(w+z+4wz)^2}/[{z(1+8w)(z^2-4w^2)(z+4w+8wz)}]$ ,
$y = 8 z(1+2z)$.
However, this change of variables leads to rather long expressions.
} in \eqref{eq:roots} is given by
\bea
x &=& \frac{(1-w)^2}{w} \, ,
\label{newchange1} \nonumber \\
y &=& \frac{1-w+w^2-z^2}{z^2-w} \, .
\label{newchange2}
\eea
Since we want to express all of the MIs in terms of $w$ and $z$, it is also necessary to write the Mandelstam variable $u$ in terms of $w$ and $z$. Starting from the relation $u = -s-t+2m^2$ one  finds
\be
\frac{u}{m^2} = \frac{w^2-(1-w+w^2)z^2}{w(w-z^2)} \, .
\ee
The analytic expressions that can be found in the ancillary files are written in terms of $w$ and $z$ defined in~(\ref{newchange2}).

In order to write $w$ and $z$ as a function of $x$ and $y$, it is necessary to invert the system in~(\ref{newchange2}).
The first equation has two solutions
\bea
w_1 &=& \frac{\sqrt{x+4}-\sqrt{x}}{\sqrt{x+4}+\sqrt{x}} \, ,
\nonumber \\
w_2 &=& \frac{\sqrt{x+4}+\sqrt{x}}{\sqrt{x+4}-\sqrt{x}} = \frac{1}{w_1} \, .
\eea
When $0 < x < \infty$ ($-\infty < s < 0$), the first solution is limited and such that $0<w_1<1$, while the second solution is unlimited, $1<w_2<\infty$. In view of the fact that GPLs of $w$ should be manifestly real in this region, we choose the first solution, $w_1$.
The second equation,~(\ref{newchange2}), has four solutions, two for each choice of $w$:
\bea
z_1 &=& \sqrt{\frac{1-w+w^2+wy}{1+y}} \, ,
\nonumber \\
z_2 &=& - \sqrt{\frac{1-w+w^2+wy}{1+y}} = - z_1 \, .
\eea
The first solution $z_1$ is always positive for $y>0$ ($t<0$) and $w=w_1$, while the second is always negative. 
When, for a given $x$, $y$ varies from 0 to $\infty$, $z_1$ is limited to the range
$\sqrt{w} < z_1 < \sqrt{1-w+w^2}$.

Consequently, in the region $s<0$ and $t<0$, we choose the set of variables
\be
w = \frac{\sqrt{x+4}-\sqrt{x}}{\sqrt{x+4}+\sqrt{x}} \, , \qquad
z = \sqrt{\frac{1-w+w^2+wy}{1+y}} \, .
\label{wzvar}
\ee

When $s$ becomes positive, it is necessary to consider the Feynman prescription and add to $s$ a positive vanishing imaginary part: $s+i0^+$:
\be
x = - \frac{s+i0^+}{m^2} \equiv - x' - i0^+ \, ,
\ee
where, now, $x'=s/m^2>0$. If $x'$ is such that $0<x'<4$, $w$ becomes a phase and
it moves on the upper unit circle:
\be
w = \frac{\sqrt{4-x'}-\sqrt{-x'-i0^+}}{\sqrt{4-x'}+\sqrt{-x'-i0^+}} =  
\frac{\sqrt{4-x'}+i\sqrt{x'}}{\sqrt{4-x'}-i\sqrt{x'}} = e^{i 2\phi} \, ,
\label{wb0and4}
\ee
with
\be
\phi = \arctan \left( \sqrt{\frac{x'}{4-x'}} \right) \, .
\ee
For $x'=4$, $w$ becomes real again and one finds that $w=-1$. 

For physical kinematics one finds that $s>4m^2$, $t<0$, $u<0$, where
\be
t_{min} < t < t_{max} \, ,
\ee
with
\bea
t_{min} & = & m^2-\frac{s}{2} - \frac{1}{2} \sqrt{s(s-4m^2)}  \, , \\
t_{max} & = & m^2-\frac{s}{2} + \frac{1}{2} \sqrt{s(s-4m^2)} \, .
\eea
In this physical region we use
\be
w = \frac{\sqrt{x+4}-\sqrt{x}}{\sqrt{x+4}+\sqrt{x}}
 = \frac{\sqrt{x'-4}-\sqrt{x'}}{\sqrt{x'-4}+\sqrt{x'}} + i 0^+ \, , \qquad
z = \sqrt{\frac{1-w+w^2+wy}{1+y}} \, .
\label{wzvar}
\ee
with the phase space constraint
\bea
0< -w < z < 1\,.
\label{wzrange}
\eea
The crossing $t \leftrightarrow u$ is given by
\bea
z \to -\frac{w}{z}\,.
\eea
By keeping into account the relation \eqref{wzrange},
the roots in \eqref{eq:roots} become 
\bea
\sqrt{s(s-4m^2)} &=&  m^2 \frac{w^2 - 1}{w} \nonumber \\
\sqrt{m^2 s (m^2-t)(s+t-m^2)} &=& m^4 \frac{(w -1 )^3 z}{w (z^2 - w)}\,.
\label{root2rational}
\eea
We apply these substitutions to the definition of the canonical integrals in the physical
region in equations (\ref{eq:4.1} -- \ref{eq:4.96}) and employ the resulting rational expressions to define the Canonical
Basis also in other regions of the phase space.

\section{Integration and Results}
\label{sec:integration}

In terms of the rationalizing variables, we can write for the matrix
$\tilde{A}(\vec{x})$
in our differential equation \eqref{CanSys}
\bea
\tilde{A}(\vec{x}) = \sum_k \tilde{A}^{(k)} \ln(l_k)
\eea
where the $\tilde{A}^{(k)}$ are rational matrices and the letters
$l_k$ form the alphabet
\begin{align}
\{ l_k \} =  \Big\{ &   w, w - 1, w + 1, z, z-1, z+1, w - z, w + z, w - z^2, w^2 - w  + 1 - z^2,
\notag \\
&  w^2 - z^2(w^2 -w + 1) , w^2 -3w+ z^2 + 1 \Big\}.
\end{align}
The last letter is needed only for Topology B.
We provide explicit expressions for the matrix $\tilde{A}(\vec{x})$ for Topologies $A$ and $B$, respectively, in the ancillary files on {\tt arXiv}.

This alphabet allows to analytically integrate the MIs in terms of GPLs of
argument $w$
with the weights
\begin{align}
 \Biggl\{ & 0, 1, -1, z, -z, z^2, \frac{1 - \sqrt{4 z^2-3}}{2}, \frac{1 + \sqrt{4 z^2-3}}{2}, 
  \frac{z(z - \sqrt{4 - 3z^2})}{2(z^2-1)}, \notag\\
  & \frac{z(z + \sqrt{4 - 3z^2})}{2(z^2-1)}, 
  \frac{3 - \sqrt{5 - 4 z^2}}{2}, \frac{3 + \sqrt{5 - 4 z^2}}{2}  \Biggr\} \, ,
\end{align}
and GPLs of argument $z$ with the weights
\be
\{ 0,-1,1,-i,i \} \, .
\ee
We fix the boundary constants by imposing regularity conditions, supplemented by
external input for a few well-known simple integrals.

The analytic continuation of the GPL functions of $w$ and $z$ between different regions of the phase space is non-trivial. We found it convenient to provide the MIs in terms of an analytic expression which is valid in the region $s<0$
and of a second analytic expression that is valid in the region $s>0$.
Our complete results are available in the ancillary files
{\tt sol-A-unphys.m}, {\tt sol-B-unphys.m} for $s<0$ and
{\tt sol-A-phys.m}, {\tt sol-B-phys.m} for $s>0$.

\section{Numerical Checks}
\label{numericalchecks}
In order to validate our results we performed numerical checks in different points of the phase space. In several cases, the MIs were checked by evaluating the MIs in the  pre-canonical basis numerically by means of Sector Decomposition \cite{Binoth:2000ps} as implemented in {\tt SecDec}
\cite{Carter:2010hi,Borowka:2012yc,Borowka:2015mxa,Borowka:2017idc} and {\tt FIESTA}  \cite{Smirnov:2008py,Smirnov:2013eza,Smirnov:2015mct} and by subsequently comparing the numerical results with the evluation of the analytic expressions for the MIs carried out with {\tt GiNaC} \cite{Bauer:2000cp}.

However, in some cases, and in particular for the MIs involving six or seven denominators,
we were not able to obtain sufficiently precise numbers by a direct evaluation of the MIs with {\tt SecDec} or {\tt FIESTA}.
For this reason, we employed the techniques described in  \cite{vonManteuffel:2014qoa,Panzer:2014gra,vonManteuffel:2017myy} and rewrote the canonical MIs
as linear combinations of quasi-finite integrals. Quasi-finite integrals are integrals which have, at worst, a single pole in $\epsilon$ which originates from the Euler Gamma function prefactor in the Feynman parameter representation of the integral. Quasi-finite integrals are built with the same set of propagators as the original integral but they might be defined in shifted space-time dimensions and might have one or more propagators squared or raised to higher power. It was shown in \cite{vonManteuffel:2014qoa, vonManteuffel:2017myy} that quasi-finite integrals are evaluated more efficiently by {\tt SecDec} and {\tt FIESTA} with respect to non quasi-finite integrals with the same sets of propagators.
Using {\tt SecDec} we generated numerical results for quasi-finite integrals in the unphysical and in the physical region.
Subsequently, we converted these numbers to results for the canonical integrals,
at which level we were left with typically 2-6 significant digits, depending on the integral
and the region of phase space.
We successfully compared these numbers against those obtained from the analytic expressions of the MIs, which are the main result of the present work. With this procedure it was possible to test numerically all of the 52+44 MIs evaluated in this paper.

In addition, we compared numerically the MIs that are in common with the ones presented in \cite{DiVita:2018nnh}
with a numeric evaluation of their expressions, finding complete agreement.
Finally, all of the MIs evaluated in this work were simultaneously evaluated in \cite{PD}. We compared numerically the MIs evaluated in this work with the results obtained in \cite{PD}. This comparison was carried out in several phase space points, both in the physical region ($s > 4 m^2$) and in the non-physical region ($s<0$). We found complete agreement between the results in this work and the ones in \cite{PD}.

\section{Conclusions}
\label{conclusions}
In this paper we presented the analytic calculation of the master integrals necessary for the evaluation of the last two color coefficients of the interference between two-loop and tree-level diagrams for the partonic process $q\bar{q} \to t\bar{t}$, for which an analytic expression is not yet available.

The master integrals were evaluated with the method of differential equations. By determining a canonical basis, we brought the system of first-order linear differential equations into an $\epsilon$-form, allowing for their decoupling after an expansion in powers of the dimensional regulator $\epsilon$. We integrated the expansion coefficients in terms of generalized harmonic polylogarithms of two dimensionless variables through to weight 4 and fixed the integration constants using regularity conditions and known solutions for simple integrals.

We checked our analytic results numerically against the results obtained with two numerical codes, {\tt SecDec} and {\tt FIESTA}, using the method of quasi-finite integrals.
We also compared numerically the MIs that are in common with the ones presented in \cite{DiVita:2018nnh}, finding complete agreement. Finally, we cross-checked numerically the MIs of our topology A with the authors of \cite{PD}, in several points of the phase space, finding complete agreement.

All the analytic expressions for the MIs presented in this paper are provided as computer readable ancillary files together with the arXiv submission of this work.

\section{Acknowledgments}

We are grateful to L.~Tancredi for early collaborations and for providing the rationalizing parametrization used in this work. We would like to thank S.~Di~Vita, T.~Gehrmann, S.~Laporta, P.~Mastrolia, A.~Primo, and U.~Schubert for the cross checks of the MIs presented in this work, in several points of the phase space.
The work of A.F.\ is supported in part by the National Science Foundation under Grant No.~PHY-1417354 and PSC CUNY Research Award TRADA-61151-00 49.
The work of A.v.M.\ is supported in part by the National Science Foundation under Grant No.~1719863.
Feynman diagrams are drawn with Axodraw \cite{Vermaseren:1994je}. 

\appendix

\section{Pre-canonical form for the Master Integrals}
\label{pre-can}

In this section we present the routing for all the pre-canonical MIs of the two
topologies.  The definition of the integration measure ${\mathcal D}^dk_i$ ($i=1,2$) can be found in \eqref{eq:intmeas}. The list of the denominators $D_i$ ($i =1, \cdots, 9$) can be found in \eqref{eq:dens}. 

\subsection{Topology $A$}

\begin{align}
\T^{A}_{1}&=\intd\frac{1}{D_{8}^2 D_{9}^2}, \Space\quad
\T^{A}_{2}=\intd\frac{1}{D_5^2 D_8^2 D_9} , \\
\T^A_3&=\intd\frac{1}{D_5 D_8^2 D_9^2},\Space
\T^A_4=\intd\frac{1}{D_5^2 D_7 D^2_9} , \\
\T^A_5&=\intd\frac{1}{D_3^2 D_7 D^2_9}, \Space
\T^A_6=\intd\frac{1}{D_3^2 D_7^2 D_9} , \\
\T^A_7&=\intd\frac{1}{D_2^2 D_3 D^8}, \Space
\T^A_8=\intd\frac{1}{D_2^2 D_3^2 D_8} , \\
\T^A_9&=\intd\frac{1}{D_1^2 D_5 D_9^2},\Space
\T^A_{10}=\intd\frac{1}{D_1^2 D_2 D_7^2} , \\
\T^A_{11}&=\intd\frac{1}{D_5 D_7 D_8 D_9^2},\qquad\qquad\;\;
\T^A_{12}=\intd\frac{1}{D_3 D_7 D_8 D_9^2} , \\
\T^A_{13}&=\intd\frac{1}{D_3 D_7 D_8 D_9^3}, \qquad\qquad\;\;
\T^A_{14}=\intd\frac{1}{D_2 D_3 D_8^2 D_9} , \\
\T^A_{15}&=\intd\frac{1}{D_2 D_3 D_8^3 D_9},\, \qquad\qquad\;\;
\T^A_{16}=\intd\frac{1}{D_1 D_5 D_7 D^2_9} , \\
\T^A_{17}&=\intd\frac{1}{D_1 D_5 D_7 D^3_9},\, \qquad\qquad\;\;
\T^A_{18}=\intd\frac{1}{D_1 D_5 D_7^2 D^2_9} , \\
\T^A_{19}&=\intd\frac{1}{D_1^2 D_2 D_7 D_9},\, \qquad\qquad\;\,
\T^A_{20}=\intd\frac{1}{D_3 D_5 D_7 D_8 D_9} , \\
\T^A_{21}&=\intd\frac{1}{D_3 D_5 D_7 D_8 D_9^2},\qquad\quad\,
\T^A_{22}=\intd\frac{1}{D_3 D_5 D_7 D_8^2 D_9} , \\
\T^A_{23}&=\intd\frac{1}{D_2 D_3 D_7 D_8 D_9},\qquad\quad\,
\T^A_{24}=\intd\frac{1}{D_2 D_3 D_7 D_8 D_9^2} , \\
\T^A_{25}&=\intd\frac{1}{D_2 D_3^2 D_7 D_8 D_9},\qquad\quad\,
\T^A_{26}=\intd\frac{1}{D_2 D_3 D_7 D_8^2 D_9} , \\
\T^A_{27}&=\intd\frac{1}{D_2 D_3 D_5 D_8 D_9},\qquad\quad\,
\T^A_{28}=\intd\frac{1}{D_2 D_3 D_5 D_8 D_9^2} , \\
\T^A_{29}&=\intd\frac{1}{D_2 D_3 D_5 D_8^2 D_9},\qquad\quad\,
\T^A_{30}=\intd\frac{1}{D_1 D_5 D_7 D_8 D_9} , \\
\T^A_{31}&=\intd\frac{1}{D_1 D_5 D_7 D_8 D_9^2},\qquad\quad\,
\T^A_{32}=\intd\frac{1}{D_1 D_3 D_5 D_7 D_9^2} , \\
\T^A_{33}&=\intd\frac{1}{D_1 D_3 D_5 D_7^2 D_9},\qquad\quad\,
\T^A_{34}=\intd\frac{1}{D_1 D_2 D_7 D_8 D_9} , \\
\T^A_{35}&=\intd\frac{1}{D_1 D_2 D_7^2 D_8 D_9},\qquad\quad\,
\T^A_{36}=\intd\frac{1}{D_1 D_2 D_3 D_7 D_9} , \\
\T^A_{37}&=\intd\frac{1}{D_1 D_2 D_3 D_7 D_9^2},\qquad\quad\,
\T^A_{38}=\intd\frac{1}{D_1 D_2 D_3 D_7 D_8} , \\
\T^A_{39}&=\intd\frac{1}{D_1 D_2 D_3 D_7 D_8^2},\qquad\quad\,
\T^A_{40}=\intd\frac{1}{D_1 D_2 D_3 D_5 D_8^2} , \\
\T^A_{41}&=\intd\frac{1}{D_1 D_2 D_3 D_5 D_8^3},\qquad\quad\,
\T^A_{42}=\intd\frac{1}{D_1 D_3 D_7 D_8 D_9} , \\
\T^A_{43}&=\intd\frac{1}{D_1 D_2 D_5 D_7 D_8 D_9},\quad\quad
\T^A_{44}=\intd\frac{1}{D_1 D_2 D_3 D_7 D_8 D_9} , \\
\T^A_{45}&=\intd\frac{D_4}{D_1 D_2 D_3 D_7 D_8 D_9},\quad\quad
\T^A_{46}=\intd\frac{D_5}{D_1 D_2 D_3 D_7 D_8 D_9} , \\
\T^A_{47}&=\intd\frac{D_6}{D_1 D_2 D_3 D_7 D_8 D_9},\quad\quad
\T^A_{48}=\intd\frac{1}{D_1 D_2 D_3 D_5 D_8 D_9} , \\
\T^A_{49}&=\intd\frac{1}{D_1 D_2 D_3 D_5 D_7 D_8 D_9},\;\;\;
\T^A_{50}=\intd\frac{D_4}{D_1 D_2 D_3 D_5 D_7 D_8 D_9} , \\
\T^A_{51}&=\intd\frac{D_6}{D_1 D_2 D_3 D_5 D_7 D_8 D_9},\;\;\;
\T^A_{52}=\intd\frac{D_4 D_6}{D_1 D_2 D_3 D_5 D_7 D_8 D_9} .
\end{align}

\subsection{Topology $B$}

\begin{align}
\T^{B}_{1}&=\intd\frac{1}{D_{8}^2 D_{9}^2}, \quad\Space 
\T^{B}_{2}=\intd\frac{1}{D_{4}D_{8}^2 D_{9}^2} , \\
\T^{B}_{3}&=\intd\frac{1}{D_{3}^2 D_{7} D_{9}^2}, \Space
\T^{B}_{4}=\intd\frac{1}{D_{3}^2 D_{7}^2 D_{9}} , \\
\T^{B}_{5}&=\intd\frac{1}{D_{3}^2 D_{4} D_{9}^2}, \Space
\T^{B}_{6}=\intd\frac{1}{D_{2}^2 D_{3} D_{8}^2} , \\
\T^{B}_{7}&=\intd\frac{1}{D_{2}^2 D_{3}^2 D_{8}},\Space
\T^{B}_{8}=\intd\frac{1}{D_{1}^2 D_{2}^2 D_{7}} , \\
\T^{B}_{9}&=\intd\frac{1}{D_{3}D_{7}D_{8}D_{9}^2},\qquad\qquad
\T^{B}_{10}=\intd\frac{1}{D_{3}D_{7}D_{8}D_{9}^3} , \\
\T^{B}_{11}&=\intd\frac{1}{D_{3}D_{4}D_{8}D_{9}^2},\qquad\qquad
\T^{B}_{12}=\intd\frac{1}{D_{3}D_{4}D_{8}D_{9}^3} , \\
\T^{B}_{13}&=\intd\frac{1}{D_{3}D_{4}D_{8}^2 D_{9}^2},\qquad\qquad
\T^{B}_{14}=\intd\frac{1}{D_{2}D_{3}D_{8}^2 D_{9}} , \\
\T^{B}_{15}&=\intd\frac{1}{D_{2}D_{3}D_{8}^3 D_{9}},\qquad\qquad
\T^{B}_{16}=\intd\frac{1}{D_{1}^2 D_{2}D_{7}D_{9}} , \\
\T^{B}_{17}&=\intd\frac{1}{D_{1}D_{2}^2 D_{4} D_{8}}, \qquad\quad\;\;\;
\T^{B}_{18}=\intd\frac{1}{D_{3}D_{4}D_{7}D_{8}D_{9}^2} , \\
\T^{B}_{19}&=\intd\frac{1}{D_{3}D_{4}D_{7}D_{8}D_{9}^3}, \qquad\quad\;
\T^{B}_{20}=\intd\frac{1}{D_{2}D_{3}D_{7}D_{8}D_{9}} , \\
\T^{B}_{21}&=\intd\frac{1}{D_{2}D_{3}D_{7}D_{8}D_{9}^2},\qquad\quad\;
\T^{B}_{22}=\intd\frac{1}{D_{2}D_{3}^2 D_{7}D_{8}D_{9}} , \\
\T^{B}_{23}&=\intd\frac{1}{D_{2}D_{3}D_{7}D_{8}^2D_{9}},\qquad\quad\;
\T^{B}_{24}=\intd\frac{1}{D_{1}D_{3}D_{4}D_{7}D_{9}} , \\
\T^{B}_{25}&=\intd\frac{1}{D_{1}D_{3}D_{4}D_{7}D_{9}^2},\qquad\quad\;
\T^{B}_{26}=\intd\frac{1}{D_{1}D_{2}D_{7}^2 D_{8}D_{9}} , \\
\T^{B}_{27}&=\intd\frac{1}{D_{1}D_{2}D_{7} D_{8}D_{9}},\qquad\quad\;
\T^{B}_{28}=\intd\frac{1}{D_{1}D_{2}D_{4}D_{8}D_{9}} , \\
\T^{B}_{29}&=\intd\frac{1}{D_{1}D_{2}D_{4}D_{8}^2 D_{9}},\qquad\quad\;
\T^{B}_{30}=\intd\frac{1}{D_{1}D_{2}D_{4}D_{7}D_{9}} , \\
\T^{B}_{31}&=\intd\frac{1}{D_{1}D_{2}D_{4}D_{8}D_{9}^2},\qquad\quad\;
\T^{B}_{32}=\intd\frac{1}{D_{1}D_{2}^2 D_{4} D_{7}D_{8}} , \\
\T^{B}_{33}&=\intd\frac{1}{D_{1}D_{2}D_{3}D_{7}D_{9}}, \qquad\quad\;
\T^{B}_{34}=\intd\frac{1}{D_{1}D_{2}D_{3}D_{7}D_{9}^2} , \\
\T^{B}_{35}&=\intd\frac{1}{D_{1}D_{2}D_{3}D_{4}D_{8}D_{9}},\qquad
\T^{B}_{36}=\intd\frac{1}{D_{1}D_{2}D_{4}D_{7}D_{8}D_{9}} , \\
\T^{B}_{37}&=\intd\frac{D_{6}}{D_{1}D_{2}D_{4}D_{7}D_{8}D_{9}},\qquad
\T^{B}_{38}=\intd\frac{1}{D_{1}D_{2}D_{3}D_{7}D_{8}D_{9}} , \\
\T^{B}_{39}&=\intd\frac{D_{4}}{D_{1}D_{2}D_{3}D_{7}D_{8}D_{9}},\qquad
\T^{B}_{40}=\intd\frac{D_{5}}{D_{1}D_{2}D_{3}D_{7}D_{8}D_{9}} , \\
\T^{B}_{41}&=\intd\frac{D_{6}}{D_{1}D_{2}D_{3}D_{7}D_{8}D_{9}},\qquad
\T^{B}_{42}=\intd\frac{1}{D_{1}D_{2}D_{3}D_{4}D_{7}D_{8}D_{9}} , \\
\T^{B}_{43}&=\intd\frac{D_{5}}{D_{1}D_{2}D_{3}D_{4}D_{7}D_{8}D_{9}},\;\;\,
\T^{B}_{44}=\intd\frac{D_{6}}{D_{1}D_{2}D_{3}D_{4}D_{7}D_{8}D_{9}} .
\end{align}

\section{Numerical Results}
\label{num-res}

In this Appendix we collect numerical results for the seven-denominator canonical MIs at the point 
\begin{equation}
m=1 \, \text{GeV} \, , \qquad s = 5.1 \, \text{GeV}^2  \, , \quad \text{and} \quad t = - 2.5\, \text{GeV}^2  \, .
\end{equation}

The numerical values of the MIs are (with 16 significant digits):

\begin{align}
f^A_{49} =& - 0.8125 \nn\\
          & + (1.571461643987763 - i\,1.570796326794896 ) \epsilon \nn\\
          & + (1.869800565465933 + i\,7.871341877028778 ) \epsilon^2  \nn\\
          & - (26.64417846013623 + i\,2.934819494524318 ) \epsilon^3  \nn\\
	  & - (5.561888073241050 + i\,69.90392666348392 ) \epsilon^4 \, , \\
f^A_{50} =& \phantom{+} (0.3936751877201319 - i\,1.229555494857724 ) \epsilon^2  \nn\\
          & + (14.12478202913410 - i\,2.239408800880071 ) \epsilon^3  \nn\\
	  & + (49.29394916594301 + i\,37.08333857464637 ) \epsilon^4 \, , \\
f^A_{51} =& \phantom{+}  0.02083333333333333  \nn\\
          & + 0.07833393820762243 \epsilon  \nn\\
	  & + (8.538951737141223 - i\,1.580009353612773 ) \epsilon^2 \nn\\
	  & - (4.529079554851615 + i\,2.930479163733208 ) \epsilon^3 \nn\\
	  & + (0.1103747892867767 - i\,78.51623866876891 ) \epsilon^4 \, , \\
f^A_{52} =& - 0.0625  \nn\\
          & + 0.2937522682785850 \epsilon  \nn\\
	  & + (7.8274169758047892 + i\,5.478308035237822 ) \epsilon^2 \nn\\
	  & - (26.357322954146530 - i\,15.39070197526472 ) \epsilon^3 \nn\\
	  & - (121.01714343276939 + i\,42.90574414612206 ) \epsilon^4 \, , \\
f^B_{42} =& - (0.5193031088754503 + i\,0.7853981633974483 ) \epsilon  \nn\\
          & - (5.247646105592740 - i\,3.6418501617483698 ) \epsilon^2 \nn\\
	  & - (51.07989282173662 + i\,30.039485666638732 ) \epsilon^3 \nn\\
	  & - (68.03046563599218 + i\,107.21203451885746 ) \epsilon^4 \, , \\
f^B_{43} =& - 1  \nn\\
          & + (2.702244138720994 - i\,3.9269908169872423 ) \epsilon  \nn\\
	  & + (18.05310915519800 + i\,12.796025276368288 ) \epsilon^2  \nn\\
	  & - (3.231845611282520 + i\,2.7724176443956750 ) \epsilon^3  \nn\\
	  & + (127.3934689436371 - i\,12.984632048850981 ) \epsilon^4 \, , \\
f^B_{44} =& - 0.625  \nn\\
          & + (0.9285563596344188 + i\,3.1415926535897928 ) \epsilon  \nn\\
	  & - (6.716934387387509 + i\,10.089909832711628 ) \epsilon^2  \nn\\
	  & + (50.38595267312016 - i\,44.149878496215228 ) \epsilon^3  \nn\\
	  & + (146.4134579642496 - i\,86.512460126974730 ) \epsilon^4 \, . 
\end{align}

\bibliographystyle{JHEP}
\bibliography{biblio}

\end{document}